%% file: main.tex
\documentclass[10pt,journal,compsoc]{IEEEtran}

\usepackage{multirow}
\usepackage{tabularx,booktabs}
\newcolumntype{Y}{>{\centering\arraybackslash}X}
\usepackage{amssymb}

\usepackage[dvipsnames]{xcolor}

\ifCLASSOPTIONcompsoc
  \usepackage[nocompress]{cite}
\else
  \usepackage{cite}
\fi

\ifCLASSINFOpdf
  \usepackage[pdftex]{graphicx}
\else
\fi

\usepackage{amsmath}
\usepackage{amsfonts}
\DeclareMathOperator*{\argmax}{arg\,max}

\usepackage{pgfplots}
\pgfplotsset{width=\linewidth,compat=1.9}

\begin{document}

\title{Demystifying Membership Inference Attacks in Machine Learning as a Service}
%
%
%
%

\author{Stacey Truex,
        Ling Liu,
        Mehmet Emre Gursoy,
        Lei Yu,
        and Wenqi Wei
\IEEEcompsocitemizethanks{\IEEEcompsocthanksitem S.Truex, L. Liu, M.E. Gursoy, L. Yu, and W. Wei are with the School of Computer Science, Georgia Institute of Technology, Atlanta, GA, 30332.
Email: \{staceytruex, memregursoy, leiyu, wenqiwei\}@gatech.edu, ling.liu@cc.gatech.edu}
\thanks{Manuscript received X; revised Y.}}

%
%

\markboth{IEEE Transactions on Services Computing,~Vol.~X, No.~Y, MONTH~20XX}%
{}
\IEEEtitleabstractindextext{%
\begin{abstract}
Membership inference attacks seek to infer membership of individual training instances of a model to which an adversary has black-box access through a machine learning-as-a-service API. In providing an in-depth characterization of membership privacy risks against machine learning models, this paper presents a comprehensive study towards demystifying membership inference attacks from two complimentary perspectives. First, we provide a generalized formulation of the development of a black-box membership inference attack model. Second, we characterize the importance of model choice on model vulnerability through a systematic evaluation of a variety of machine learning models and model combinations using multiple datasets. Through formal analysis and empirical evidence from extensive experimentation, we characterize under what conditions a model may be vulnerable to such black-box membership inference attacks. We show that membership inference vulnerability is data-driven and corresponding attack models are largely transferable. Though different model types display different vulnerabilities to membership inference, so do different datasets.  Our empirical results additionally show that (1) using the type of target model under attack within the attack model may not increase attack effectiveness and (2) collaborative learning exposes vulnerabilities to membership inference risks when the adversary is a participant. We also discuss countermeasure and mitigation strategies. 
\end{abstract}

}

\maketitle

\IEEEdisplaynontitleabstractindextext

%
\IEEEpeerreviewmaketitle

\IEEEraisesectionheading{\section{Introduction}\label{sec:introduction}}

%
%
%
%
\IEEEPARstart{M}{achine} learning-as-a-service has seen an explosion of interest with the development of cloud platform services. Amazon~\cite{amazonml}, Microsoft~\cite{Copeland:2015:MAP:2840185}, IBM~\cite{ibmml}, and Google~\cite{googleml} have all launched such machine learning-as-a-service platforms. These services allow companies to leverage powerful machine learning and artificial intelligence technologies without requiring in-house domain expertise. Machine learning-as-a-service platforms allow users to upload their data, run various data analytics or model building processes, and deploy trained models to services of their own. Given this landscape, new interest has been given to the potential vulnerabilities of such machine learning services.

One such vulnerability is membership inference. Let us consider a cancer treatment center with a large database of valuable patient data. Let our cancer treatment center then leverage a machine learning-as-a-service platform to develop a predictive model which, when given a patient's data as input, can predict cancer-related health outcomes. The treatment center then utilizes the cloud deployment option to create a service of their own wherein users can log in, provide their own health information, and receive predictions in return. A membership inference attack considers a scenario wherein a user of such a black-box prediction service is an adversary. This adversary can provide the health information of another individual $X$ and, based on the model's output, try to infer if $X$ is a cancer patient at the treatment center.

There are two primary parties who are interested in protecting against such membership inference attacks: patient $X$ and the cancer treatment center. Previous patients of the cancer treatment center, such as patient $X$, consider their membership private and do not want their patronage to be public knowledge. For example, consider the case of a treatment center patient, Alice. Let Alice be under consideration for a job at Bob's company. Bob can leverage the cancer treatment center's service to infer whether or not Alice is a patient. Upon learning of Alice's inclusion in the cancer treatment center's database, Bob decides not to hire Alice in favor of a candidate who, he believes, will have lower healthcare costs for the company.

In addition to concerns of patients such as Alice, we also must consider the interests of the cancer treatment center, the owner of the training dataset and the trained model under attack. In today's market, across many domains, data is considered an organizational asset~\cite{lake2013data}. While internal data has always been a driver of decision making for most companies, the role of data has been moving steadily closer to the core of many industries. A company's data therefore holds intrinsic value to the organization.
Additionally, this training data is the source of the machine learning model under attack. Training this model not only requires front-end capital, time, and resources but also holds competitive business value for the treatment center. The treatment center may even charge a fee per evaluation. It is therefore essential from the cancer treatment center's perspective to protect their private database.

{\bf Membership Inference Risks v.s. Differential Privacy.\/}
Membership inference violates the privacy of both the individual participants involved in the model training and the owner of the training dataset. The former involves membership privacy of individuals who are participants in the model training and the latter involves risks of unauthorized leakages of business value or trade secrets. The ultimate goal of membership privacy is to protect against the risk of membership leakage of individuals in the data used in training a machine learning model.

When a model is secured by differential privacy, there exists some formal guarantee that the model trained on the original dataset $D$ will produce statistically similar predictions as a model trained on $D'$, which differs from $D$ by exactly one instance~\cite{dwork2014algorithmic}. Differential privacy therefore protects the content privacy and the output privacy of the model, often with some cost to model accuracy. Membership privacy refers to membership inference against machine learning models and is centered on inferring the membership of the input data, but not the content of input data, from the output result of the model.

{\bf Membership Inference v.s. Adversarial Examples.\/}
Adversarial learning to-date has been focused on attacking deployed deep learning models. Most existing membership inference attacks similarly attack deep learning models, utilizing deep neural networks (DNNs) for training both the target model under attack and the attack model~\cite{shokri2017membership, long2018understanding, hayes2017logan, carlini2018secret}. However, membership inference attacks are different from adversarial examples with respect to both the attack generation process and the adverse effects of successful attacks and therefore represent a separate class of security and privacy intrusion problems under the general umbrella of adversarial machine learning.
Concretely, adversarial deep learning research to date has been centered on the generation of adversarial examples by injecting the minimal amount of perturbation to a benign example required to cause a pre-trained classification model to misclassify the example with high probability. Thus, adversarial example-based attacks aim at altering the output of the model prediction without being visually noticed. On the contrary, a membership inference attack does not alter the prediction output at all and succeeds by simply making membership inference on the prediction output. 

{\bf Scope and Contributions of the paper.\/}
In this paper, we investigate membership inference attacks under a black-box access scenario in which an adversary may only probe the prediction API with input and receive the corresponding prediction output from the privately trained model. Our research results are novel from three perspectives. First, we describe a systematic approach to construct a membership inference attack model with a general formulation of each component of the attack model generation framework. We show that generating a membership inference attack model is a complex and multi-step strategic process. 
Second, to understand \textit{when} and \textit{how} membership inference attacks work and \textit{why} certain models and datasets are more vulnerable, we take a holistic approach with extensive empirical evidence to study and characterize membership inference attacks across different target model types, different types of training datasets, and different combinations of model types for generating attack training datasets and attack models. Finally, we introduce and investigate a new membership threat, insider membership inference, which is launched by a member of a federated learning system against other participants in an collaborative learning environment. As federated learning systems become more popular with promises of increased accuracy and privacy, highlighting and understanding this risk is an important part of the membership inference mitigation effort.

\section{Membership Inference Attacks}\label{sec:memb_inf_attacks}

In this section, we formalize membership inference attacks against machine learning models as follows: Given an instance $\mathbf{x}$ and black-box access to a classification model $F_t$ trained on a dataset $D$, can an adversary infer with high confidence that the instance $\mathbf{x}$ was contained in $D$ at the train time of $F_t$? This definition states that membership inference focuses on the question of the membership of $\mathbf{x}$ in $D$ and \textit{not} about the contents of $\mathbf{x}$. This divergence separates membership inference from existing areas of privacy research, such as differential privacy~\cite{blum2005practical},~\cite{vaidya2014random},~\cite{dwork2008differential} or secure multiparty computation~\cite{wu2016privately},~\cite{de2017efficient},~\cite{cramer2015secure}. Also notable is that membership inference attacks are at the local level: an adversary wishes to know if a particular $\mathbf{x}$ is in $D$ and \textit{not} $D$ in its entirety.

\begin{figure}[h]
\centering
\includegraphics[width=\columnwidth]{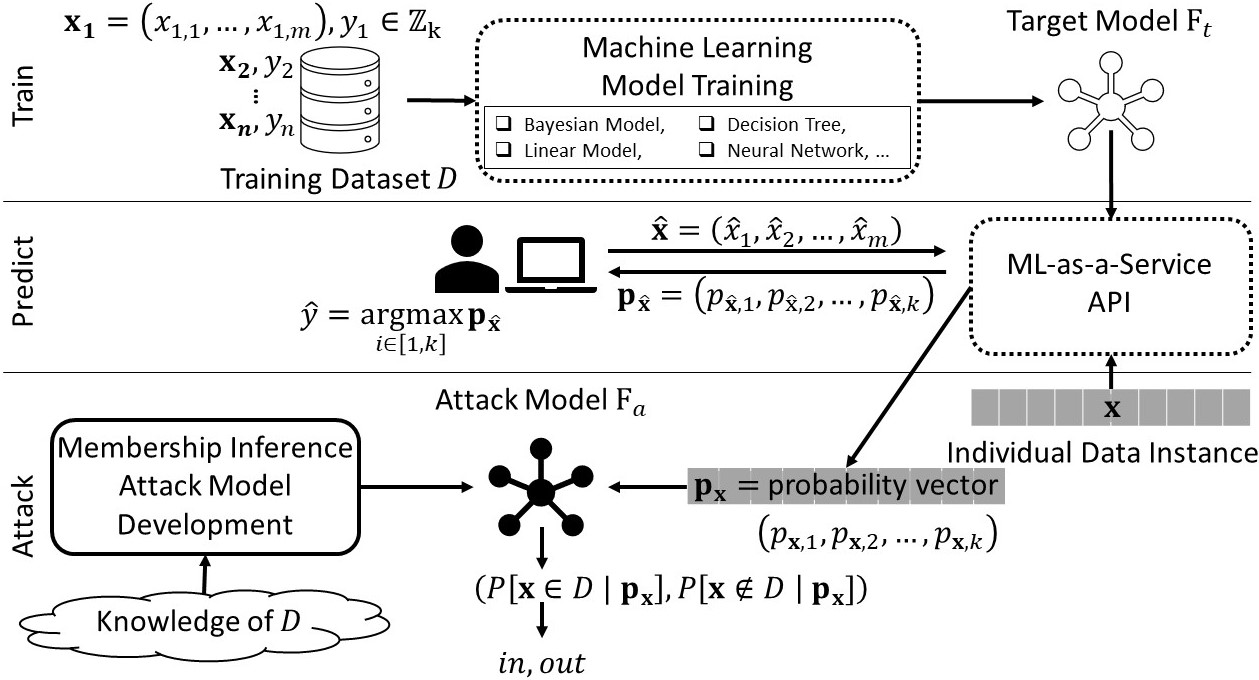}
\caption{The workflow of a Membership Inference Attack.}
\label{fig:memb_inf_attack}
\end{figure}

Figure \ref{fig:memb_inf_attack} illustrates the workflow of membership inference attack development. Given a training dataset $D$ and a classification model $F_t$ trained on $D$, the machine learning service provider may provide a classification service through a prediction API. This API offers users black-box access to the model $F_t$. Users may send prediction queries with their own data to the service and receive classification predictions. An adversary uses such a service to collect information about the private dataset $D$ on which the prediction model $F_t$ was privately trained. By leveraging any public or background knowledge of the training dataset $D$ or the target model $F_t$, an adversary builds a membership inference attack model $F_a$ to deploy for launching membership inference attacks in real time. 



To gain an in-depth understanding of the general formulation of the membership inference attack model, we first characterize the types of adversarial knowledge and datasets required to train the attack model as well as 
the attack cost, the attack value, and their evaluation metrics. We then present a systematic formulation of general attacks in Section~\ref{sec:general_model}.  

\subsection{Threat Model and Assumptions}\label{subsec:threat}

\subsubsection{Machine Learning-As-A-Service: Black-box Access}
\label{subsubsec:ml_asaservice}



Recall from Figure \ref{fig:memb_inf_attack} that the machine learning service provider publishes the trained target model $F_t$ through a black-box access API, which accepts service requests from users in the form of a prediction query for input $\mathbf{x}$ and returns the predicted class and the prediction probability vector. The input and output formats of the API are given by the service provider as part of a service agreement. However, $F_t$ and the dataset $D$ on which $F_t$ is trained remain private. Only the prediction API is exposed to users, thus ensuring only black-box access to $F_t$. 

A question one may ask is: ``How can an adversary with only black-box access to the prediction API perform membership inference attack without knowing anything about $F_t$ and the dataset $D$?'' Generally speaking, 
$F_t$ is trained as an approximation of an ideal function $y = \Gamma(\mathbf{x})$ for a training dataset $D$, where $y$ is the true class for the sample instance $\mathbf{x} \in D$. Let $y_c$ denote the output of a candidate model $F_c$.
An optimal model is then the one that minimizes the average loss defined by a chosen loss function $L(y,y_c)$ for all samples in the training set $D$, weighted by their {\em posterior} probability. The posterior probability, $P_y(y_c|\mathbf{x})$, is defined as the probability of class $y_c$ being the label of sample $\mathbf{x}$. For many application specific problems, $y = \Gamma(\mathbf{x})$ is a non-deterministic function. That is, if $\mathbf{x}$ is sampled repeatedly, different values of $y$ may be given. In this case, the optimal choice of the class for sample object $\mathbf{x}$ among all candidate class labels is the class 
that minimizes the expected loss for a given sample $\mathbf{x}$.
The target model $F_t$ is then assigned to be the optimal model for the training dataset $D$ upon the completion of the training and testing phases. 

There are currently a multitude of well-studied algorithms available to determine $F_t$. Without loss of generality to which algorithm or loss function was chosen to identify $F_t$, we simply maintain that $F_t$ is a function which maps feature vectors $\mathbf{x} \in \mathbb{R}^m$ to a class $y \in \mathbb{Z}_k$, $k\geq 2$. Our target function therefore creates a decision boundary which separates the feature space $\mathbb{R}^m$ into $k$ sets in which each set is associated with a candidate class value in $\mathbb{Z}_k$. As these sets and corresponding class assignments are chosen to minimize the loss function $L(y,y_c)$ over the training dataset $D$, the decision boundaries are strongly informed by the training dataset $D$ and will in turn be the core of the trained machine learning model $F_t$.

\subsubsection{Adversarial Knowledge}\label{subsubsec:knowledge}

We characterize the membership inference threat model based on prior adversarial knowledge. We broadly categorize this adversarial knowledge into three categories: black-box, grey-box, and white-box data knowledge. 


\textbf{Black-Box Knowledge.} An adversary is said to have black-box knowledge when the adversary does not have any \textit{specialized} knowledge of the training data. 
However, black-box knowledge may include the input and output of the service API as well as publicly available information about the target prediction model $F_t$.
For example, if the service provider is our cancer treatment center, then the adversary may have access to relevant statistics curated by the government and published for the public good including demographic information, such as the likeliness of different age groups or genders to contract certain cancers, or clinical information, such as the prevalence of co-occurrence of different diseases with various cancer types. 

\textbf{Grey-Box Knowledge.} We characterize grey-box knowledge as \textit{specialized} population-level knowledge. This may include population-level statistics that describe the distribution of features in the target model's training data. 
For example, in addition to publicly available distributions on the average age of cancer patients (black-box knowledge), the adversary may know the average age of a cancer patient seen at the target treatment center (\textit{specific} knowledge). 


\textbf{White-Box Knowledge.} White-box knowledge characterizes scenarios where the training data for $F_t$ is sampled from a constrained population or in a skewed fashion such that an adversary has access to some versions of real data in the training data $D$ of the target model $F_t$ or some leaked portion of $D$ but not the complete training set $D$. For example, a noisy version of the real data may be accessible which resembles $D$ with the addition of some noise or missing values~\cite{shokri2017membership}. Adversaries with white-box knowledge can therefore develop or access \textbf{true} ``{\tt in}'' samples and employ active learning techniques on these known samples to develop a very accurate dataset to mirror $D$. 
 

Adversaries with white-box knowledge are the most powerful whereas adversaries with only black-box knowledge represent the most difficult attack environment wherein adversaries are limited to (i) publicly availbale information, (ii) black-box queries to the prediction API, and (iii) the output of classification prediction from the target model $F_t$. This is the setting we use to formulate membership inference attacks and to characterize adverse effects and divergence of membership inferences.

\subsection{Attack Value vs. Attack Cost}\label{subsec:value_cost}

It is generally accepted that systems security should never operate as an all-or-nothing mechanism. Systems must always seek to optimize two sets of factors: cost of defense vs value of assets to the system owner and cost of attack vs value of assets to the adversary. These principles hold true to deployed learning systems with respect to membership inference attacks. 

In the context of membership inference attacks, value can be characterized by attack accuracy as an evaluation of what level of leakage is present in $F_t$ or what amount of knowledge an attacker can expect to gain. The cost of attack refers to the knowledge and work necessary for an adversary to launch a successful attack. When characterizing cost, we consider knowledge cost as well as development cost. For example, to launch an effective attack, how much knowledge of $D$ or $F_t$ does an adversary need? Gaining this knowledge should be considered a type of cost for the adversary. Alternatively, cost can also be characterized by how computationally expensive it is to develop an effective attack model $F_a$.

The accuracy of an attack can be characterized by a number of metrics. For example, an accuracy measure may be defined as the likelihood that $F_a$ correctly identifies $\mathbf{x} \in D$ \textit{or} $\mathbf{x} \notin D$. 
Alternatively, accuracy can be defined as a precision measure which indicates the fraction of the instances inferred as members that are indeed members of the training dataset $D$, focusing on the probability $Pr[\mathbf{x} \in D | F_a$ says $\mathbf{x} \in D]$.

\section{General Attack Formulation}\label{sec:general_model}

\begin{figure}[h]
\centering
\includegraphics[width=\columnwidth]{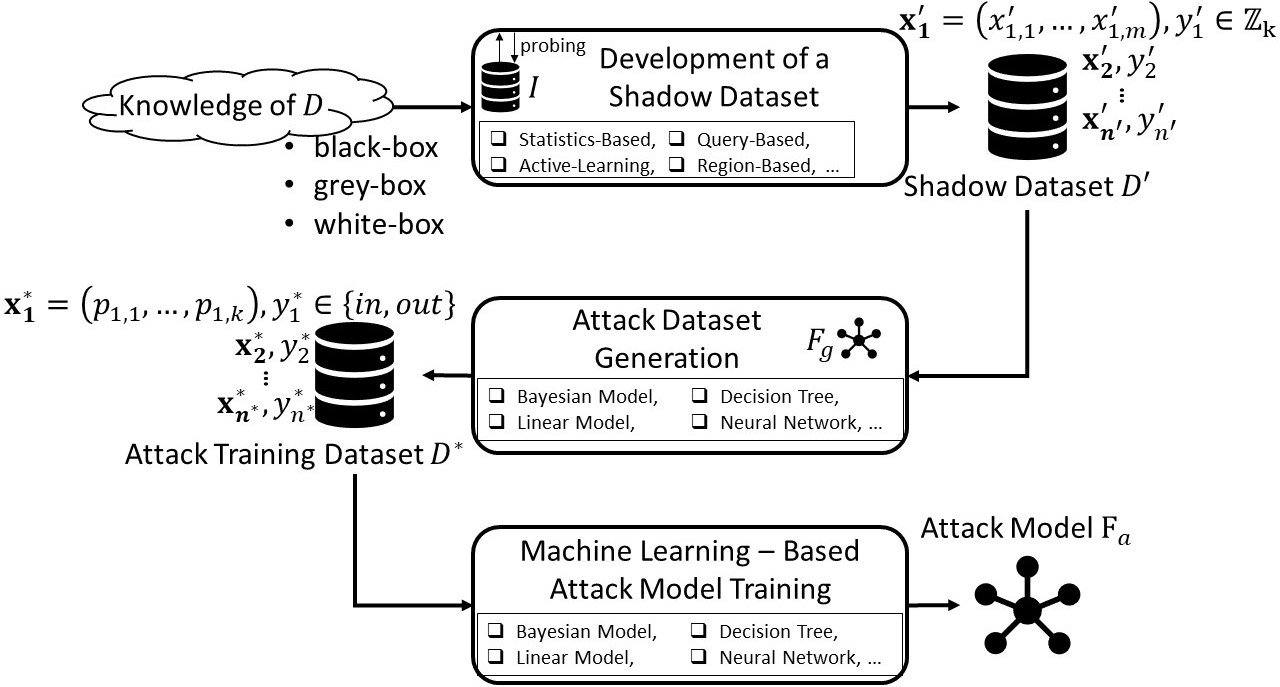}
\caption{Membership Attack Model Development.}
\label{fig:memb_inf_attack_mod_dev}
\end{figure}

Using correct identification to denote attack accuracy, the problem of membership inference is defined as follows: given a query input $\mathbf{x}$ and black-box access to the target model $F_t$, the membership inference attack answers the question of whether $\mathbf{x}\in D$ is true or false. The attack is successful if the attacker can determine with high confidence that $\mathbf{x}\in D$ is true or with high confidence that $\mathbf{x} \notin D$. 

At the most abstract level, membership inference attack models are binary classifiers. Given an instance $\mathbf{x}$ and a target model $F_t$, the goal of a membership inference attack model is to identify whether or not $\mathbf{x}$ was contained within the dataset $D$ used to train $F_t$.

Let $D$ consist of $n$ training instances $(\mathbf{x_1},y_1),(\mathbf{x_2},y_2),...,(\mathbf{x_n},y_n)$ where $\mathbf{x_i}$ consists of $m$ features, denoted by $\mathbf{x_i} = (x_{i,1},x_{i,2},...,x_{i,m})$, and  
$y_i \in \mathbb{Z}_k$, where $k$ is a finite integer value $\geq 2$.
Let $F_t:\mathbb{R}^m \rightarrow \mathbb{R}^k$ be the target model trained using this dataset $D$.
Given a particular \textit{feature vector} $\mathbf{x} \in \mathbb{R}^m$, $F_t$ will then output a probability vector $\mathbf{p} \in \mathbb{R}^k$ of the form $\mathbf{p} = (p_1,p_2,...,p_k)$, where $p_i \in [0,1] \forall i$, and $\sum_{i=1}^k p_i = 1$. The prediction class label $y$ for a feature vector $\mathbf{x}$ is the class with highest probability value in $\mathbf{p}$. Therefore $y = \argmax_{i \in \mathbb{Z}_k} F_t(\mathbf{x})$. 

Given the adversary's black-box access to $F_t$ via the prediction service API, an adversary is able to query $F_t$ with any number of instances to receive corresponding probability vectors. The adversary uses this probing access, along with any prior knowledge, to generate $I$, a representation of adversarial knowledge of $D$. 
The first building block for implementing a black-box membership inference attack is to leverage $I$ to generate a synthetic labeled dataset $D'$ to mirror the data in $D$. 
This synthetic, labeled dataset $D'$ is artificially simulated and called a {\em shadow dataset} of $D$. 
Although the word ``shadow'' was borrowed from shadow copying for systems creating back up data copies~\cite{sankaran2004volume}, the shadow dataset in our context should be thought of as a synthetic version of the real training dataset $D$.
$D'$ is then used to generate an attack training dataset $D^*$, which is required to train the final membership attack model, a binary classifier $F_a$. 

Figure \ref{fig:memb_inf_attack_mod_dev} highlights these three primary phases in the development of the membership inference attack: (1) development of a shadow dataset, (2) generation of an attack model training dataset, and (3) training and deployment of the membership inference attack model. In phase (1) the attacker's goal is to develop a dataset denoted as $D'$ which closely emulates, or \textit{shadows}, the dataset $D$ which was used to train the target model. In phase (2) the attacker uses this shadow dataset to develop a shadow model which is considered to emulate the behavior of the target model. The attacker may then observe the behavior of this shadow model in response to instances which the attacker knows were given during training versus those that were not. This behavior is used to develop an attack dataset which captures the difference between output for instances in the training data and those previously unseen by the model. Finally, in phase (3), this attack data is used to generate a binary classifier which provides predictions on whether an instance was previously known to a model based on that model's output from that instance. We now discuss each of these phases in further detail.



\subsection{Development of a Shadow Dataset}\label{subsec:develop_shadow_dataset}

\begin{figure}[h]
\centering
\includegraphics[width=\columnwidth]{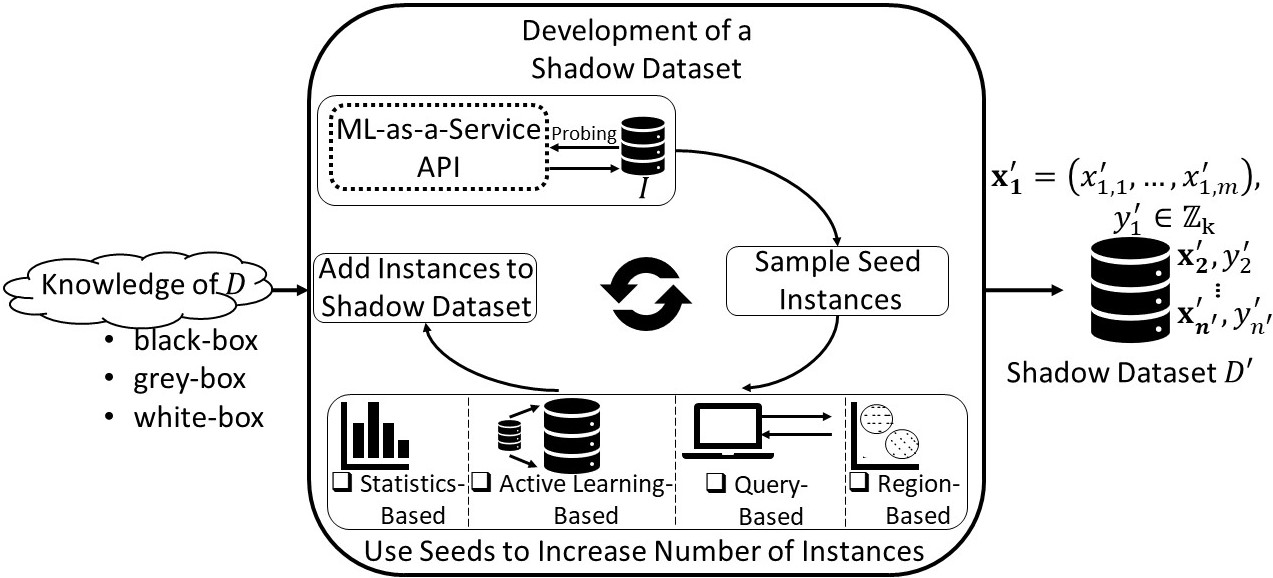}
\caption{Development of a Shadow Dataset.}
\label{fig:shadow_data_dev}
\end{figure}

Given a target model $F_t$, its training dataset $D$, and black-box adversarial knowledge 
, the development of a shadow dataset $D'$ is the first step in generating a membership inference attack model. 
$D'$ consists of $n'$ training instances $(\mathbf{x'}_1,y'_1),(\mathbf{x'}_2,y'_2),...,(\mathbf{x'_{n'}},y'_{n'})$ where each $\mathbf{x'_i}$ consists of $m$ features equivalent to those in $D$ and each $y'_i$ is a predicted class label in $\mathbb{Z}_k$. Note that $k$ and $m$ are known via the service API and thus consistent across $D$ and $D'$. The cardinality of $D$, however, remains unknown and therefore $n$ and $n'$ are likely to differ.
The shadow dataset generation process leverages the prediction service API to manage the creation and control the quality of $D'$, as shown in Figure~\ref{fig:shadow_data_dev}.

\subsubsection{API Probing}\label{subsubsec:probing}

While the training set $D$ and the cardinality (size) $n$ of $D$ are unknown to an adversary, the adversary can probe the service API to reveal structural information, such as the number of features $m$, the data types of those features, and the number of classes $k$. This knowledge can be obtained by the adversary through sending in trial query instances and observing responses.

We refer to the complete set of adversarial knowledge as $I$. This includes prior knowledge as well as that inferred from this API probing. We do not state any limitations on $I$ except that $D \not\subset I$, as the membership inference attack becomes trivial when $D\subset I$. The cost of launching an effective membership inference attack includes the work of this probing phase by the adversary.

As a result of this API probing, the adversary can construct a skeleton dataset $D'$, which is similar to $D$ in structure and, ideally, any $\mathbf{x'} \in D'$ should be a viable instance that could be included in $D$.




\subsubsection{Shadow Data Generation}\label{subsub:shadow_data_gen}
There are several ways to generate a quality shadow dataset with a small number of query probing attempts. Below we highlight four categories of techniques: statistics-based, active learning-based, query-based, and region-based generation.


\textbf{Statistics-Based Generation.} In statistics-based generation, the adversary leverages population-level statistics of the features in $D$ to create samples for $D'$. Given known distributions for features, an adversary may conduct random sampling to construct these new samples. Features may be treated independently where an instance is generated through $m+1$ random samplings of $m+1$ distributions, each distribution corresponding to either a different feature or the class label. Alternatively, sampling may account for feature relationships. This may be done, for example, when our adversary has knowledge of statistics on disease co-occurrence. 

\textbf{Active Learning-Based Generation.} Active learning is a technique developed in the semi-supervised machine learning domain \cite{zhu2005semi}. Active learning has been developed to address the problem of a largely unlabeled training dataset when assigning accurate labels is an expensive task. For example, in the development of a spam filter one may have access to a large number of unlabeled emails~\cite{sculley2007online}. It is a very expensive proposal to suggest that a human read millions of emails to provide labels, yet labels are necessary to develop an accurate filter. 
To address this problem, representative samples are selected and labeled. Given this subset of training instances which now have accurate labels, an automated process takes over and propagates the label logic to other instances. Active learning techniques may be combined with statistics-based generation in black-box or grey-box data knowledge scenarios where a large number of samples may be generated through random sampling of the features but class labels assigned through intervention by the adversary followed by active learning. 

\textbf{Query-Based Generation.} 
When using query-based generation, an adversary will generate a random sample and then query the target model. The target model will then provide a probability vector output. In this case the adversary will want to identify instances for which the machine learning service provides a class label with relatively high confidence. That is, the adversary will search for instances in which the output $\mathbf{p}$ has a $max(\mathbf{p})$ value above some predefined threshold. This, again, may be combined with other techniques. For example, query-based generation may be used to provide a seed for active learning or statistics-based generation may inform the development of the instances sent to the service provider.

\textbf{Region-Based Generation.} Region-based generation follows a clustering-based logic. Given an instance $\mathbf{x}$ with label $y$, region-based generation will seek to generate instances $\mathbf{x'}$ where $dist(\mathbf{x},\mathbf{x'})$ is below some pre-determined threshold for a pre-chosen distance function $dist$. The new instance $\mathbf{x'}$ is then assigned the same label $y$. 
One way an adversary may use region-based generation is in conjunction with white-box data knowledge. Given knowledge of some instances either in $D$ or very similar to those in $D$ an adversary can use region-based generation to expand this knowledge into a larger number of highly accurate instances to construct $D'$.

Several factors may determine which concrete technique will be chosen by an adversary, such as the knowledge contained in $I$ or the query probing results. For example, an adversary who has grey-box data knowledge may be more likely to rely on statistics-based generation due to the specificity of the available statistics to $D$ whereas an adversary with black-box knowledge may want to augment statistics-based generation with query-based generation due to lower confidence in their non-specific, population-level knowledge of $D$. On the other hand, active learning-based or region-based generation would likely be a popular choice for adversaries with white-box data knowledge where the adversary may leverage their insider knowledge of $D$. These are just some of the considerations in the development of an effective membership inference attack model.

\begin{figure}[h]
\centering
\includegraphics[width=\columnwidth]{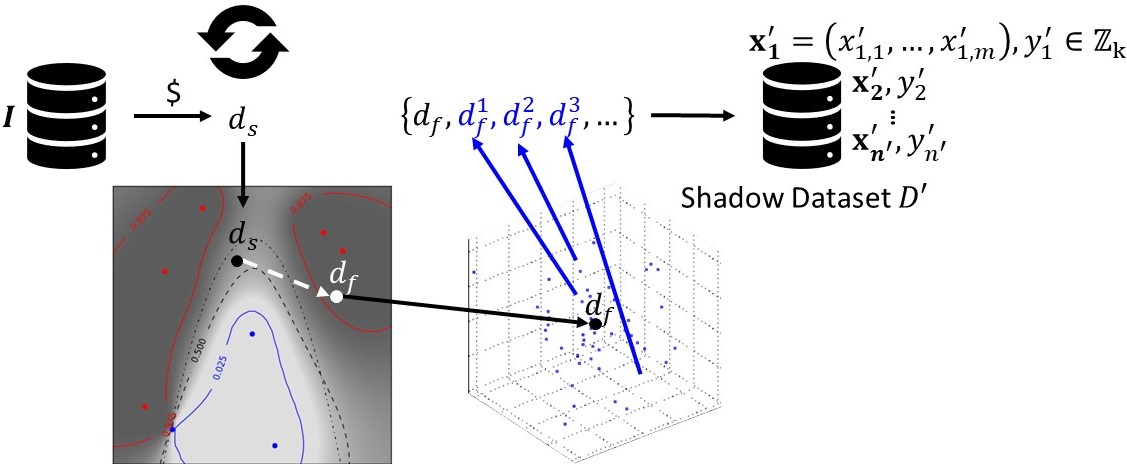}
\caption{Shadow dataset development using query-based and region-based techniques with black-box data knowledge. Figures adapted from images in~\cite{scikit-learn} and~\cite{MATLAB:2010}}
\label{fig:example_d'}
\end{figure}

In Figure \ref{fig:example_d'} we show an example of the shadow dataset development process using a combination of query-based and region-based techniques for an adversary with black-box data knowledge. The adversary first randomly generates a starting point $d_s$ using distributions in $I$. This point $d_s$ is then queried to the service provider which provides $\mathbf{p}_{d_s}$ in return. The point $d_s$ is then updated randomly and queried again. This is continued until a confidence threshold or stopping condition is reached. That is, the process stops when either (1) a point $d_f$ is found such that  
$max(\mathbf{p}_{d_f})$ meets a predefined confidence threshold or (2) the point has been updated and queried without meeting the confidence threshold the maximum number of times as determined by the attacker. If condition (2) is met a new $d_s$ is chosen until the process results in condition (1) and an accepted point $d_f$.

A hyper-cube is then constructed surrounding $d_f$. A set of new samples $\{d^1_f,d^2_f,d^3_f,...\}$ are then generated by randomly sampling from the hyper-cube region. Each point $d^i_f$ is assigned the class $\argmax_{i \in \mathbb{Z}_k} \mathbf{p}_{d_f}$ and added, along with $d_f$ to the shadow dataset $D'$. The entire process is then repeated beginning with sampling a new starting point from $I$ and ending with a new set of samples added to $D'$. The adversary will continue this repetition until a satisfactory number of samples $n'$ have been added to $D'$.

\subsection{Generation of an Attack Model Training Set}\label{subsec:generate_attack_data}

\begin{figure}[h]
\centering
\includegraphics[width=\columnwidth]{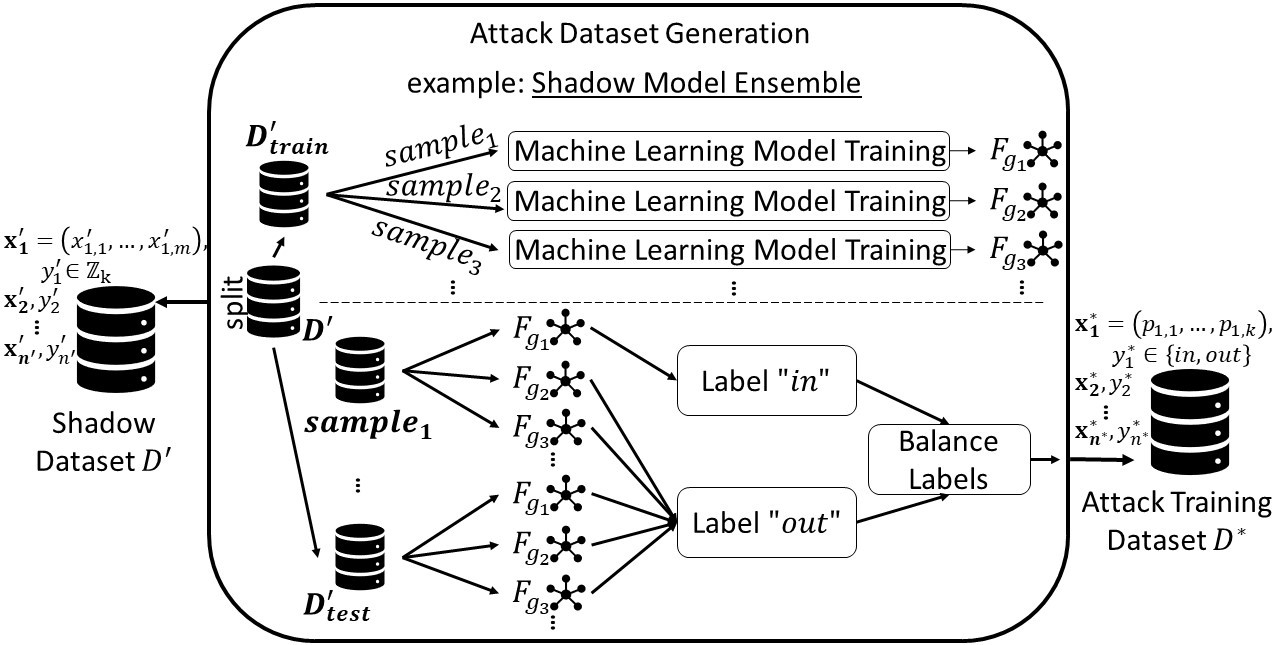}
\caption{Generation of an Attack Model Training Set.}
\label{fig:attack_data_dev}
\end{figure}

Upon the completion of the development of the shadow dataset $D'$, the adversary will proceed to utilize the shadow dataset $D'$ to develop the membership attack dataset for training a binary classifier as the final attack model, as shown in Figure~\ref{fig:attack_data_dev}. Given that each instance in $D'$ consists of a feature-vector and its known-class, denoted by $(\mathbf{x'},y')$, 
the adversary can define an attack generation function denoted by $F_g:(\mathbb{R}^m,\mathbb{Z}_k) \rightarrow (\mathbb{R}^k,\mathbb{Z}_2)$. 
$F_g$ takes a feature vector-known class pair $(\mathbf{x'},y')$ as input and outputs an attack training instance, consisting of two pieces of information: a probability vector $\mathbf{p} = (p_1,p_2,...,p_k)$ and a binary class label, indicating ``{\tt in}'' or ``{\tt out}''.  
There are several approaches to generate $F_g$ using $D'$. For example, the adversary can train a new model over $D'$ which simulates the private target model $F_t$. In this case, we call $F_g$ a shadow model of $F_t$.
Given that the adversary does not know the original training set $D$ nor the size of $D$, the adversary may leverage ensemble learning techniques~\cite{dietterich2000ensemble}, such as data partition-based ensembles, model-based ensembles, or hybrid ensemble models, to improve the quality of the shadow model with the goal of accurately simulating the target model $F_t$.
The attack model training set generation function $F_g: (\mathbb{R}^m,\mathbb{Z}_k) \rightarrow (\mathbb{R}^k,\mathbb{Z}_2)$ can therefore be viewed as an ensemble of a set of shadow models. These shadow models seek to characterize the \textit{decision boundary} of the target model. More specifically, the shadow models aim at mirroring the sensitivity of the target decision boundary to individual instances.

Consider a data partition based ensemble approach~\cite{shokri2017membership}. The adversary partitions the shadow dataset $D'$ into $D'_{train}$ and $D'_{test}$. $D'_{train}$ is then divided into $q$ partitions ($q>1$), one partition for each shadow model. Each partition of $D'_{train}$ will then be used to train a single shadow model $F_{g_i}$. Here we intentionally do not specify the machine learning model type of $F_{g_i}$ as this is yet another design choice made by an adversary. The decision may be informed if the adversary knows the model type of $F_t$ or chosen using some other criteria, we leave this decision unspecified to remove the constraint that an adversary must know the model type of $F_t$. Next, $D'_{test}$ will be evaluated against $F_{g_i}$. The corresponding outputs will then be labeled as ``{\tt out}''. Additionally, a sample of size $|D'_{test}|$ is taken from the $D'_{train}$ partition used to train $F_{g_i}$ and  evaluated against $F_{g_i}$ with the corresponding outputs labeled as ``{\tt in}''. By combining these output-label pairs, we obtain the attack train data $D^*$. 

Figure \ref{fig:attack_data_dev} highlights the workflow of generating an attack model training set. 
An adversary may choose a single model for efficiency or an ensemble of models to increase the size or generality of $D^*$. An adversary 
may diversify model types in an ensemble when $F_t$'s type is unknown. If an ensemble is used, there are choices on size, sampling, and aggregation that must be made and can be informed by the adversary's knowledge of the target in various ways. We stress that while we formulate the membership inference attack using a general model, there exist many implementation variants. 

{\bf Effect of Ensemble Methods.\/} Combining multiple different models reduces the risk of choosing the wrong hypothesis within the hypothesis space of a particular problem. Also, multiple models allows for more effective local search, which many machine learning algorithms perform in various ways, and limits the impact of the local optima problem. Finally, a combining of chosen hypotheses allows for an expansion of the hypothesis space~\cite{dietterich2000ensemble}. 
Two common ways to accomplish this diversity are bagging and boosting.

Bagging is accomplished by either drawing a sample of training examples from the original dataset randomly and with replacement, or by creating disjoint subsets of the original training data called cross-validated committees~\cite{parmanto1996improving}. 
A commonly used implementation of bagging is the Random Forest ensemble model~\cite{breiman2001random}.

Using the boosting technique, a set of weights is maintained for each instance within the training dataset. Each model is then trained iteratively to minimize the \textit{weighted} error of the training dataset. 
The weights of the training instances are updated to put more emphasis on the misclassified examples. The adaboost ensemble model~\cite{freund1995desicion} is a popular implementation of boosting. 

The use of multiple models via ensemble learning for attack data generation 
decreases the risk of choosing the wrong hypothesis.
Given an adversary who has only black-box knowledge of the target model (regardless of the adversary's knowledge type of the dataset $D$), there is no guaranteed method to reproducing the target model's behavior on $D$. A diversity of generation models will minimize the risk that the adversary is only capturing one behavior type or candidate decision boundary shape. This again accentuates the need for boosting or bagging to ensure that the model set is diverse.


\subsection{Generating the Membership Attack Model}\label{subsec:train_deploy_attack}

\begin{figure}[h]
\centering
\includegraphics[width=\columnwidth]{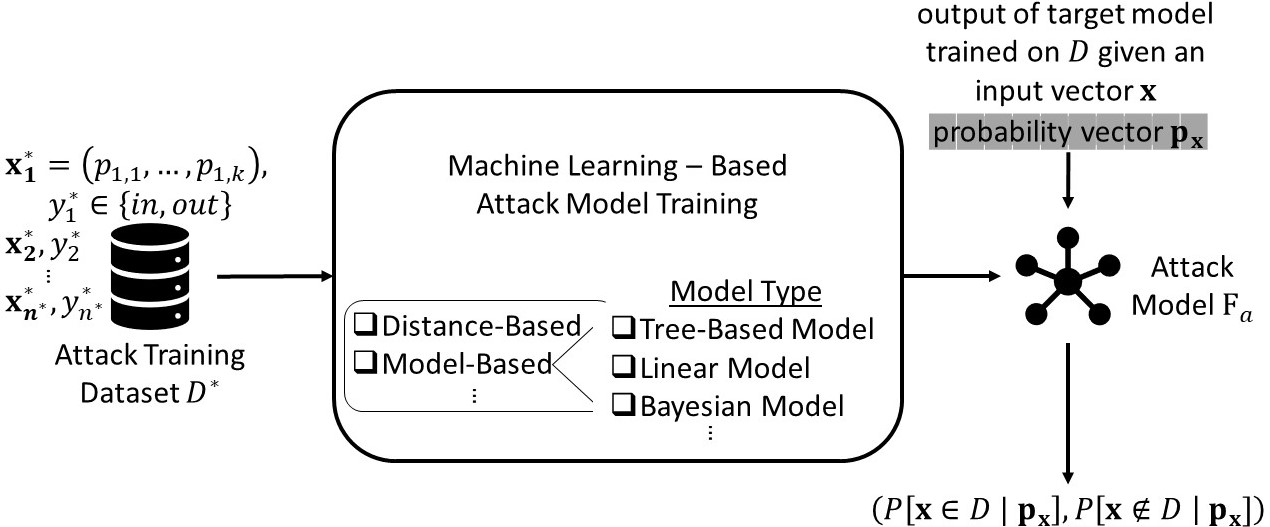}
\caption{Training and Deployment of the Membership Inference Attack Model}
\label{fig:attack_mod_train}
\end{figure}

The attack model training set $D^*$ contains the outputs from the generation function $F_g:(\mathbb{R}^m,\mathbb{Z}_k) \rightarrow (\mathbb{R}^k,\mathbb{Z}_2)$. $D^*$ consists of $n^*$ instances, $(\mathbf{x^{*}_1},y^{*}_1), (\mathbf{x^{*}_2},y^{*}_2), ..., (\mathbf{x^{*}_{n^*}},y^{*}_{n^*})$, and each instance is the output of $F_g$ for some input $(\mathbf{x'},y') \in (\mathbb{R}^m,\mathbb{Z}_k)$. This attack model training dataset $D^*$ will then be used to generate the final attack model $F_a:\mathbb{R}^k \rightarrow \{${\tt in},{\tt out}$\}$, which takes as input a probability vector output for an instance $\mathbf{x}$ and outputs a binary classification of ``{\tt in}'' or``{\tt out}''. 
Similarly, we make no assumptions on how $D^*$ is used to inform $F_a$ but rather say that $D^*$ is available to an adversary during the generation of the membership attack model $F_a$. A number of machine learning models and techniques can be leveraged to train a binary classification-based attack model $F_a$ using $D^*$. $F_a$ can then be deployed against the output of the target model $F_t$ such that, ideally, given an instance $\mathbf{x}$, $F_a(F_t(\mathbf{x})) =$ ``{\tt in}'' if the instance $\mathbf{x} \in D$ and $F_a(F_t(\mathbf{x})) =$ ``{\tt out}'' if $\mathbf{x} \notin D$. Figure \ref{fig:attack_mod_train} visualizes this final phase of developing a membership inference attack.

Regardless of how complex the chosen training process is, whether it is a distance evaluation or a complex machine learning model, this phase produces the final attack model $F_a$ which will be deployed for membership inference attack against $D$ in real time. 

\section{General Attack Characterization}\label{sec:demystify}

We have thus far provided a general formulation of membership inference attacks. In this section, we characterize such attacks through a systematic evaluation of a variety of machine learning models and model combinations using multiple datasets. We show that membership inference vulnerability is data-driven and its attack models are largely transferable. Although the target model is a dominating factor in determining vulnerability, attack data generation techniques need not explicitly mirror the target model. Finally, we show that membership inference attacks can persist as insider attacks in federated systems.


\subsection{Experimental Setup}\label{subsec:exp_setup}

We conduct a series of experiments to support our characterization of membership inference attacks. In this section we report empirical evidence for four types of machine learning models (logistic regression, k-nearest neighbor, decision tree, and Na\"ive Bayes). A total of seven datasets are used in these experiments: Adult, MNIST, CIFAR-10, Purchases-10, Purchases-20, Purchases-50, and Purchases-100. Our experiments were conducted using Python~\cite{python} and algorithms available with scikit-learn~\cite{scikit-learn}. All reported results are averaged across 10 runs each using 10-fold cross-validation.

In reporting our results we use the accuracy metric. This is to capture the attacker's ability to both identify when $\mathbf{x} \in D$ \underline{and} $\mathbf{x} \notin D$. Previous work, namely~\cite{shokri2017membership}, report the precision metric which captures an attacker's ability to correctly identify when $\mathbf{x} \in D$ only. We therefore include this metric in Table \ref{tab:modvariation} to compare the authors' neural network results with those from our experiments with logistic regression, k-nearest neighbor, decision tree, and Na\"ive Bayes models. Precision is also reported, along side accuracy, in Table \ref{tab:insider_acc} to aide in the comparison between insider and outsider attacks. All other results report the accuracy metric.


\subsubsection{Datasets}

\paragraph{Adult}

The Adult dataset is available on the UCI Machine Learning Repository~\cite{Dua:2017} and contains 48,842 instances described by 14 different features. The feature set contains both continuous (ex: age, hours per week) and discrete (ex: education, marital status) values. This dataset presents a binary classification problem wherein one wishes to identify if an individual makes $\geq \$50$K or $< \$50$K in yearly salary. 

\paragraph{MNIST}

MNIST is a publicly available dataset containing 70,000 images of handwritten digits~\cite{lecun2010mnist}. Each image is formatted to be 32 x 32 and processed such that the digit is at the center of the image. The MNIST dataset constitutes a 10-class classification problem where the task is to identify which digit between $0$ and $9$, inclusive, is contained within a given image.

\paragraph{CIFAR-10}

The CIFAR-10 dataset, also publicly available, contains 60,000 color images~\cite{krizhevsky2009learning}. Again, each image is formatted to be 32 x 32. The CIFAR-10 dataset also has 10 classes: airplane, automobile, bird, cat, deer, dog, frog, horse, ship and truck. Each class has 6,000 available images. The problem is therefore a 10-class classification problem where the task is to identify which of the 10 classes is depicted in a given image.

\paragraph{Purchases}

Finally, we developed a number of purchases datasets similar to what was done in~\cite{shokri2017membership}. The purchases datasets were developed from the Kaggle Acquire Valued Shoppers Challenge dataset which contains the shopping history of several thousand individuals. From this dataset we create new datasets wherein each instance represents an individual and each feature represents a particular product. If an individual has purchased this product, there will be a 1 for the feature and otherwise a 0. The instances are then clustered into different shopping profile types. These cluster assignments are treated as the classes. We created different datasets with 10, 20, 50, and 100 shopping profile types. The classification problem then becomes: given a shopper's purchase history, identify their shopping profile type.

\medbreak

We highlight these datasets as each shows fundamentally different results in previous work and the makeup of each is significantly different. For example, the Adult dataset presents a binary classification problem with numeric and factor features and was demonstrated to be resilient to membership inference attacks in previous work with precision results of 50.3\%. By comparison, the Purchases-50 dataset has 50 different class values, only binary features, and was much more vulnerable to attack in previous work, with precision results of 86.0\%~\cite{shokri2017membership}.


\subsection{Membership Inference: Data-Driven}\label{subsec:data_driven}

Recall Section \ref{sec:general_model}. There are many choices an adversary can make during the three phases of developing the membership inference attack model, each of which may impact the model accuracy and consequently the attack success rate. Two elements that the adversary cannot control, however, are the training dataset $D$ and the target model $F_t$. 
Our experimental results show that membership inference attacks are data-driven. That is, the make-up of $D$ strongly correlates with the corresponding target model $F_t$'s vulnerability to membership inference attacks. Potential targets of such attacks can therefore use knowledge of their data to evaluate their risk.
Table \ref{tab:data_compare} compares the seven different datasets measuring three characteristics: (1) feature distribution, (2) number of classes (size of $k$), and (3) accuracy of (vulnerability to) membership inference.

\begin{table}[h]
    \centering
    \resizebox{\columnwidth}{!}{\begin{tabular}{|c||c|c|c|}
        \hline
        Dataset         & In-Class Standard Deviation   & Number of Classes & Accuracy of       \\
                        &                               &                   & Membership Inference  \\
        \hline
        \hline
        Adult           & 0.1433                        & 2                 & 59.89                 \\
        MNIST           & 0.1586                        & 10                & 61.75                 \\
        CIFAR-10        & 0.2301                        & 10                & 90.44                 \\
        Purchases-10    & 0.3820                        & 10                & 82.29                 \\
        Purchases-20    & 0.3873                        & 20                & 88.98                 \\
        Purchases-50    & 0.3873                        & 50                & 93.71                 \\
        Purchases-100   & 0.3832                        & 100               & 95.74                 \\  
        \hline
    \end{tabular}}\newline
    \caption{\small{Comparison of datasets versus membership inference attack accuracy using a decision tree model.}}
    \label{tab:data_compare}
\end{table}


The number of classes is important as it characterizes the number of regions into which the input space $\mathbb{R}^m$ is divided. The more classes, the smaller each region. With smaller regions, there is less uninformed space and in fact the regions will more tightly surround the provided training instances in $D$. This will make any single instance more likely to alter the decision boundary as space is ``tighter'' between the regions.
If an instance is more likely to impact the decision boundary of the target model, then an adversary will be more likely to infer its inclusion in the training dataset.

Another side of this same argument can be seen in the in-class standard deviation metric. This value captures feature distributions by addressing the following: within a dataset, given all instances of the same class, how similar are the feature vectors? The Adult and MNIST datasets, for example, have significantly lower standard deviations than the Purchases datasets despite having more instances of each class. This demonstrates more uniformity within classes of the Adult and MNIST datasets as compared to the Purchases datasets. If an instance is exceedingly similar to other instances of the same class then it will be less likely to noticeably impact the decision boundary during training. If, however, instances within a class are notably different, then the inclusion of each instance may significantly impact the decision boundary. Therefore, the uniformity of the target training data within each class will, in addition to the number of classes, impact an adversary's ability to identify the inclusion of a particular instance.

The variation in attack accuracy results reported in Table \ref{tab:data_compare} from experiments using the same attack development process targeting decision tree models trained on each of the seven datasets demonstrates that factors in the dataset such as the in-class standard deviation and the number of classes 
contribute to a model's susceptibility to membership inference. We make two interesting observations. First, the impact of in-class standard deviation varies for different types of datasets and different scales of $k$-class classification problems. Second, membership inference vulnerability is related to a number of factors, many of which are tied to the dataset itself. Concretely, for image datasets the in-class standard deviation increases as the complexity of the dataset increases. For example, MNIST represents images of grey scale for 10 single digits and CIFAR-10 represents color images of 10 different complex entities ranging from flowers to animals. It is observed that the attacks have higher success rate for more complex images which also show larger in-class SD. On the other hand, consider the Purchase datasets where the value of $k$ varies from 2 to 10, 20, 50 and, finally, 100. The attack accuracy increases as $k$ increases without the standard deviation increasing. This shows, from a different perspective, that attack accuracy is higher for more complex classification problems. For example, consider the Purchases-50 and Purchases-100 datasets. The increase in attack accuracy demonstrates that in-class standard deviation may not always be a dominant factor in membership inference vulnerability, other data-driven factors, such as the complexity of the classification problem and the complexity of the dataset itself, can also be important factors in membership inference vulnerability.

Table \ref{tab:data_compare} shows more explicitly the impact on the attack accuracy of two parameters: the variation of in-class standard deviation and the parameter $k$ for the $k$-class classification problem. While we want to highlight the impact of these important parameters, neither the in-class standard deviation nor the value of $k$ is the sole dominating factor. Rather, the results in Table \ref{tab:data_compare} demonstrate that the membership inference vulnerability is not isolated to the learning process itself but also highly related to several attributes of the training dataset. 
This leads us to characterize membership inference attacks as data-driven. 

\subsection{Transferable Attack Models}\label{subsec:transferable}

Recent studies on adversarial machine learning attacks such as evasion attacks or poisoning attacks~\cite{szegedy2013intriguing},~\cite{papernot2016practical},~\cite{papernot2016transferability},~\cite{rozsa2016accuracy} have shown that maliciously generated adversarial examples tend to transfer from one model to another. This is an important property as it opens the door to adversaries in the black-box scenario. That is, an adversary need not know details of their target model to launch a successful attack. Through extensive experiments, we observe that membership inference attacks are similarly transferable.


For example, the results in Table \ref{tab:purch20_carttarget} are the accuracy of membership inference attacks using various attack configurations against a decision tree model trained on the Purchases-20 dataset. The relative consistency seen in Table \ref{tab:purch20_carttarget} demonstrates that many attack configurations are viable. An adversary does not need to know the target model configuration to launch an effective attack.
This suggests that membership inference attack models are transferable from one target model to another, provided that targets are trained on the same dataset $D$.



\begin{table}[h]
    \centering
    \begin{tabular}{|c||c|c|c|c|}
        \hline
        \textbf{Purchases-20}   & \multicolumn{4}{|c|}{Attack Data Generation Model} \\
        \hline
        Attack Model            & DT                & k-NN              & LR                & NB    \\
        \hline
        DT                      & \textbf{88.98}    & \textbf{87.49}    & 72.08             & 81.84 \\
        k-NN                    & \textbf{88.23}    & 72.57             & \textbf{84.75}    & 74.27 \\
        LR                      & \textbf{89.02}    & \textbf{88.11}    & \textbf{88.99}    & 83.57 \\
        NB                      & \textbf{88.96}    & 78.60             & \textbf{89.05}    & 66.34 \\
        \hline
    \end{tabular}\newline
    \caption{\small{Accuracy of membership inference attack against a decision tree target model trained on the Purchases-20 dataset.}} 
    \label{tab:purch20_carttarget}
\end{table}

Next, we evaluate the standard deviation of membership inference attack results for a membership inference attack model $F_a$ learned over the output of an attack data generation model $F_g$ and deployed against a target model $F_t$ across the seven datasets for the following three scenarios: (1) vary the model types for $F_g$ and $F_a$ while keeping $F_t$ consistent, (2) varying $F_t$ and $F_a$ while keeping $F_g$ consistent, and (3) varying $F_t$ and $F_g$ while keeping $F_a$ consistent. 

Table \ref{tab:full_cifar_10_acc} shows the results of the experiments on various combinations of model types for the CIFAR-10 dataset. We calculate the average standard deviation for scenario (1) as follows: $DT_{sd}$ denotes the standard deviation of all accuracy values in rows 1, 5, 9, and 13 where the target model is a decision tree (DT). Similarly, $k$-$NN_{sd}$ corresponds to rows 2, 6, 10, and 14, $LR_{sd}$ with 3, 7, 11, and 15, and $NB_{sd}$ with 4, 8, 12, and 16. Then we consider the average standard deviation in accuracy for scenario (1) to be the average of $DT_{sd}$, $k$-$NN_{sd}$, $LR_{sd}$, $NB_{st}$. For scenario (3) we follow a similar process but with row sets 1-4, 5-8, 9-12, and 13-16. Scenario (2) is calculated by averaging the standard deviations for each of the 4 $F_g$ columns.

\begin{table}[h]
    \centering
    \resizebox{\columnwidth}{!}{
    \begin{tabular}{|c|c|cccc|}
        \hline
        \multirow{2}{*}{Attack Model}   & \multirow{2}{*}{Target Model} & \multicolumn{4}{c|}{Attack Data Generation Model} \\
        \cline{3-6}
                                        &                               & DT                & k-NN              & LR                & NB                \\
        \hline
        \multirow{4}{*}{DT}             & DT                            & \textbf{90.44}\%  & 85.64\%           & 60.48\%           & 65.78\%           \\
                                        & k-NN                          & 54.92\%           & 69.32\%           & 55.01\%           & 51.38\%           \\
                                        & LR                            & 53.84\%           & 61.06\%           & 61.10\%           & 50.02\%           \\
                                        & NB                            & 50.46\%           & 50.58\%           & 49.98\%           & 50.20\%           \\
        \hline
        \multirow{4}{*}{k-NN}           & DT                            & \textbf{89.96}\%  & 81.55\%           & 89.07\%           & 61.10\%           \\
                                        & k-NN                          & 55.33\%           & 68.32\%           & 62.45\%           & 50.89\%           \\
                                        & LR                            & 51.34\%           & 59.58\%           & 64.78\%           & 50.09\%           \\
                                        & NB                            & 50.12\%           & 50.61\%           & 50.46\%           & 50.11\%           \\
        \hline
        \multirow{4}{*}{LR}             & DT                            & \textbf{90.37}\%  & \textbf{90.11}\%  & 88.81\%           & \textbf{66.98}\%  \\
                                        & k-NN                          & 51.72\%           & 69.90\%           & 65.29\%           & 55.64\%           \\
                                        & LR                            & 50.01\%           & 64.34\%           & 67.40\%           & 54.49\%           \\
                                        & NB                            & 50.54\%           & 50.63\%           & 50.60\%           & 50.29\%           \\
        \hline
        \multirow{4}{*}{NB}             & DT                            & \textbf{90.42}\%  & 89.86\%           & \textbf{90.52}\%  & 63.71\%           \\
                                        & k-NN                          & 50.33\%           & 68.31\%           & 57.65\%           & 53.08\%           \\
                                        & LR                            & 50.00\%           & 64.22\%           & 67.63\%           & 53.54\%           \\
                                        & NB                            & 50.58\%           & 50.44\%           & 50.58\%           & 50.01\%           \\
        \hline
    \end{tabular}}\newline
    \caption{\small{Accuracy for CIFAR-10 dataset across experiments with various attack, data generation, and target models.}} 
    \label{tab:full_cifar_10_acc}
\end{table}

We follow this process for each of the seven datasets and summarize the results in Table \ref{tab:stdev_fixed_models}. This allows us to compare the impact of model variation for each $F_t$, $F_g$, and $F_a$. We observe that the standard deviation of membership inference attack results is relatively small against a fixed target model when compared to a fixed attack data generation model or fixed attack model. A smaller standard deviation is indicative of a larger impact. That is, the accuracy is stable when the standard deviation is small. A small standard deviation would indicate that the fixed model has more influence over attack accuracy than the varied models. 

\begin{table}[h]
    \centering
    \begin{tabularx}{\columnwidth}{ |c|| *{3}{Y|} }
        \hline
        \multirow{2}{*}{Dataset}    & \multicolumn{3}{|c|}{Standard Deviation in Accuracy Results}  \\
        \cline{2-4}
                                    & Fixed $F_t$       & Fixed $F_g$   & Fixed $F_a$                   \\
        \hline
        \hline
        Adult                       & \textbf{0.0093}   & 0.0335        & 0.0328                        \\
        MNIST                       & \textbf{0.0126}   & 0.0347        & 0.0351                        \\
        CIFAR-10                    & \textbf{0.0643}   & 0.1233        & 0.1366                        \\
        Purchases-10                & \textbf{0.0396}   & 0.1069        & 0.1074                        \\
        Purchases-20                & \textbf{0.0545}   & 0.1336        & 0.1352                        \\
        Purchases-50                & \textbf{0.0705}   & 0.1468        & 0.1482                        \\
        Purchases-100               & \textbf{0.0849}   & 0.1468        & 0.1452                        \\
        \hline
    \end{tabularx}\newline
    \caption{\small{Standard deviation between accuracy results with (1) fixed $F_t$ type and varying $F_g$ and $F_a$ types, (2) fixed $F_g$ type and varying $F_t$ and $F_a$ types, and (3) fixed $F_a$ type and varying $F_t$ and $F_g$ types.}}
    \label{tab:stdev_fixed_models}
\end{table}

Table \ref{tab:stdev_fixed_models} clearly shows that for all datasets the deviation is minimized when $F_t$ is fixed. This indicates that variation in $F_g$ and $F_a$ have a comparatively low impact on attack success rates. This supports the evidence in Table \ref{tab:purch20_carttarget} that an adversary need not have particularly informed choices in $F_g$ or $F_a$ to develop an attack model. We therefore characterize membership inference attacks, like others in the adversarial learning domain, as transferable.

In summary, Table \ref{tab:purch20_carttarget} highlights the role of the attack data generation model using the Purchases-20 dataset. It articulates that a variety of attack model structures may be used against a single target model. Tables \ref{tab:full_cifar_10_acc} and \ref{tab:stdev_fixed_models} cover combinations of target models, attack models, and attack data generation models with the CIFAR-10 dataset to illustrate a broader, more comprehensive picture that both the target model and the target dataset are the most important factors for membership inference vulnerability. For each of the four attack models (DT, k-NN, LR, NB), the results show the impact in varying target models on attack accuracy under one of the four given attack data generation models. Table \ref{tab:stdev_fixed_models} shows the derivation of accuracy results for \textit{each} of the seven datasets under a fixed target, a fixed generation model, or a fixed attack model. These results indicate that an adversary may be able to develop an attack model without knowing ``best’’ attack model or the ``best’’ attack data generation model. To further understand this observation, we next study the impact of different choices of attack model on attack accuracy. 



\subsection{Attacks Across Model Types}\label{subsec:attack_ft_type}
Many works in adversarial machine learning focus on the vulnerability of deep learning models whose uses range 
from image classification~\cite{he2016deep},~\cite{kisavcanin2017deep} to speech recognition~\cite{deng2013new},~\cite{hinton2012deep} to natural language processing~\cite{mikolov2013efficient}. 
By generating adversarial examples which are tweaked in ways that are unnoticeable to humans, adversaries can exploit the models' complexities to force misclassification~\cite{kurakin2016adversarial},~\cite{carlini2017towards}. This trend has also influenced the study of membership inference problems.


Most existing efforts on membership inference~\cite{long2018understanding},~\cite{hayes2017logan},~\cite{carlini2018secret},~\cite{shokri2017membership} have been focused on deep learning models. We argue that
the complexity exploited in traditional adversarial learning attacks is not explicitly leveraged in membership inference attacks. Additionally, areas where membership inference would be most alarming, such as healthcare~\cite{lee2017medical}, e-commerce, banking, and government often deploy simpler model types, such as decision trees as model understanding is prioritized for safe use of predictive services. This leads to a natural question: are model types outside of deep learning methods susceptible to membership inference? Our empirical study on all seven datasets and four types of models demonstrates that not only are other model types vulnerable to membership inference attacks but that the model type is in fact very influential in determining the extent of that vulnerability.

The basic hypothesis of membership inference attack is it that models respond differently to instances which they have ``seen'' versus those they have not. In this hypothesis it is clear that model behavior and sensitivity are likely to impact vulnerability. We therefore consider a variety of model types outside of neural networks. We consider models from 4 different major categories: linear models (logistic regression), Bayesian models (Na\"ive Bayes), cluster models (k-nearest neighbor with k=5), and tree models (CART decision trees). 

For equal comparison to the previous work in~\cite{shokri2017membership}, we use the shadow model implementation of the membership inference attack. All results from our experiments use 10-fold cross-validation and are averaged across 10 runs for a randomly selected sample of 10,000 instances. 


\begin{table}[h]
    \centering
    \begin{tabular}{|c|c|c|c|c|c|c|}
        \hline
        Dataset         & LR                & k-NN              & DT                & NB                & NN                \\
        \hline
        Adult           & \textit{50.13}    & 51.39             & \textbf{55.49}    & 50.22             & 50.30             \\
        MNIST           & 53.25             & \textit{50.44}    & \textbf{56.66}    & 50.48             & 51.70             \\
        CIFAR-10        & 70.25             & 65.99             & \textbf{83.94}    & \textit{50.03}    & 78.00             \\
        Purchases-10    & 64.56             & 53.53             & \textbf{73.85}    & \textit{50.61}    & 55.00             \\
        Purchases-20    & 75.85             & 55.36             & \textbf{81.94}    & \textit{50.79}    & 59.00             \\
        Purchases-50    & 81.61             & 58.19             & \textbf{88.88}    & \textit{52.08}    & 86.00             \\
        Purchases-100   & 83.78             & 60.11             & 92.19             & \textit{54.93}    & \textbf{93.50}    \\
        \hline
    \end{tabular}\newline
    \caption{\small{Precision of membership inference attack across 5 model types.}} 
    \label{tab:modvariation}
\end{table}

\begin{table}[h]
    \centering
    \begin{tabular}{|c|c|c|c|c|c|}
        \hline
        Dataset         & LR                & k-NN  & DT                & NB                \\
        \hline
        Adult           & \textit{50.17}    & 51.22 & \textbf{59.89}    & 50.18             \\
        MNIST           & 54.38             & 50.59 & \textbf{61.75}    & \textit{50.81}    \\
        CIFAR-10        & 67.40             & 68.32 & \textbf{90.37}    & \textit{50.01}    \\
        Purchases-10    & 66.82             & 53.78 & \textbf{82.29}    & \textit{51.00}    \\
        Purchases-20    & 80.50             & 55.92 & \textbf{88.98}    & \textit{51.29}    \\ 
        Purchases-50    & 88.60             & 59.57 & \textbf{93.71}    & \textit{53.49}    \\
        Purchases-100   & 90.23             & 62.19 & \textbf{95.74}    & \textit{57.61}    \\
        \hline
    \end{tabular}\newline
    \caption{\small{Accuracy of membership inference attack across 4 model types.}} 
    \label{tab:modvariation_acc}
\end{table}

We can clearly see in Table \ref{tab:modvariation} that other models are in fact vulnerable to membership inference and that both the training data (as discussed in Section \ref{subsec:data_driven}) and the model type play an important role in understanding a particular model's risk. 
As shown in Table \ref{tab:modvariation}, despite the variety in our datasets, the highest precision is seen with the decision tree model for all datasets except Purchases-100 while the Na\"ive Bayes models consistently show exceedingly low precision across all datasets.

In general, a target model whose decision boundary is unlikely to be drastically impacted by a particular instance will be more resilient to membership inference attacks. For example, the Na\"ive Bayes algorithm independently considers the probability of a given class for each feature. Therefore, given significant training samples, a single instance only marginally affects these probabilities. This explains the low numbers continuously seen when attacking a Na\"ive Bayes model. By contrast, a decision tree leaf node will consider a unique feature combination to determine class rather than each feature in isolation. The introduction of a single instance, if that instance displays a unique feature set-class combination, may cause a decision tree to grow and entire new branch. This sensitivity to single instances makes membership inference attacks more successful when targeting decision tree models.

Consequently, it is important to understand that while different datasets display different vulnerabilities to membership inference, so do different model types. 
This also indicates that machine learning-as-a-service providers who are wary of membership inference attacks against their deployed models may also be able to use model choice to help mitigate vulnerability.

We have studied the impact of various target models on attack accuracy (Section \ref{subsec:train_deploy_attack}) and the impact of various attack models on attack accuracy (Section \ref{subsec:attack_ft_type}), we next study the impact of the attack generation models on the accuracy of the membership inference attack.



\subsection{Variation in Generation Model}\label{subsec:gen_mod_var}
\begin{table*}[t]
    \centering
    \resizebox{\textwidth}{!}{
    \begin{tabular}{|c||cc|cc|cc|cc|}
        \hline
        \multirow{2}{*}{Dataset}    & Model Types for                       & Accuracy
                                    & \multirow{2}{*}{$type(F_t^{max})$}    & Accuracy  
                                    & \multirow{2}{*}{$type(F_g^{max})$}    & Accuracy          
                                    & \multirow{2}{*}{$type(F_a^{max})$}    & Accuracy              \\
                                    & $(F_t^{max}, F_g^{max}, F_a^{max})$   & $(F_t^{max}, F_g^{max}, F_a^{max})$                               
                                    &                                       & All $type(F_t^{max})$
                                    &                                       & All $type(F_g^{max})$
                                    &                                       & All $type(F_a^{max})$ \\
        \hline
        \hline
        Adult                       & (DT, DT, NB)                          & 59.91\%   
                                    & DT                                    & 59.89\%   
                                    & DT                                    & 59.89\%   
                                    & NB                                    & 50.18\%               \\
        MNIST                       & (DT, DT, LR)                          & 61.80\%   
                                    & DT                                    & 61.75\%   
                                    & DT                                    & 61.75\%   
                                    & LR                                    & 54.38\%               \\
        CIFAR-10                    & (\textbf{DT}, \textbf{LR}, NB)        & 90.52\%   
                                    & \textbf{DT}                           & 90.44\%   
                                    & \textbf{LR}                           & 67.40\%   
                                    & NB                                    & 50.01\%               \\
        Purchases-10                & (\textbf{DT}, \textbf{k-NN}, DT)      & 82.45\%   
                                    & \textbf{DT}                           & 82.29\%   
                                    & \textbf{k-NN}                         & 53.78\%   
                                    & DT                                    & 82.29\%               \\
        Purchases-20                & (\textbf{DT}, \textbf{LR}, NB)        & 89.05\%   
                                    & \textbf{DT}                           & 88.98\%   
                                    & \textbf{LR}                           & 80.50\%   
                                    & NB                                    & 51.29\%               \\
        Purchases-50                & (\textbf{DT}, \textbf{LR}, LR)        & 93.77\%   
                                    & \textbf{DT}                           & 93.71\%   
                                    & \textbf{LR}                           & 88.60\%   
                                    & LR                                    & 88.60\%               \\
        Purchases-100               & (\textbf{k-NN}, \textbf{LR}, DT)      & 95.86\%   
                                    & \textbf{k-NN}                         & 95.74\%   
                                    & \textbf{LR}                           & 90.23\%   
                                    & DT                                    & 62.19\%               \\
        \hline
    \end{tabular}}\newline
    \caption{\small{Model set up with maximum accuracy averaged across 10 runs using 10-fold cross validation. Maximum configuration is then compared to configurations where model type is consistent across the target, generation, and attack models using each model type represented in the maximum configuration.}}
    \label{tab:highest_acc_combo}
\end{table*}

Instinctually, one may believe that the model used to generate attack data must be of the same type as the target model. We briefly discuss why previous research has made this same assumption and then investigate its veracity and seek to explain why it does not strictly hold. 

In~\cite{shokri2017membership}, the authors claim that the shadow model implementation of a membership inference attack requires the shadow models be trained in a similar way to the target model, an assumption followed by later work such as~\cite{long2017towards}. Consistent with our formalization of membership inference attacks, this claim is equivalent to saying that the attack data generation technique $F_g$ must mirror the behavior of the target model $F_t$ under attack.

The reasoning behind this assumption is intuitive. Let us say the target model $F_t$ is an approximation of an ideal function $\Gamma$ for a dataset $D$. Then, given a data point $\mathbf{x}$ for which the adversary aims to identify membership in the training data $D$, the output provided to the adversary will be $F_t(\mathbf{x})$. This output will then be provided to the attack model to determine classification of $\mathbf{x}$ as ``{\tt in}'' or ``{\tt out}''. Let
the binary classifier which serves as the attack model be trained from the output of a generation model $F_g$. That is, the output $F_g(\mathbf{x'})$ for all $\mathbf{x'} \in D'$ make up the attack model training data $D^*$.

We can clearly see that our attack model is therefore trained on the output of a function $F_g$ and deployed against the output of a function $F_t$. It is understandable then that previous work would seek to have the behavior of $F_g$ mirror the behavior of $F_t$. This assumption naturally extends to say that $F_g$, in an attempt to mirror $F_t$, must be, or intuitively should be, of the same model type as $F_t$ in a successful membership inference attack model.

However, this assumption is not necessary to launch an effective membership inference attack. 
We conducted a set of experiments varying model type combinations of $F_t$, $F_g$, and $F_a$ considering the four candidate model types.
The combination with the highest accuracy for each dataset is reported in Table \ref{tab:highest_acc_combo}. Let $type(F_t^{max}), type(F_g^{max}), type(F_a^{max})$ be the target model, attack data generation model, and attack model types respectively for the membership inference attack which reported the highest accuracy. These types will naturally vary for different datasets. We also report accuracy for scenarios when all three model types are set to $type(F_t^{max})$. We similarly report when all three types are equivalent  to $type(F_g^{max})$ and $type(F_a^{max})$. 

For example, we recall that in Table \ref{tab:full_cifar_10_acc} the highest accuracy was reported when the target model was a decision tree, the data generation model was a logistic regression model, and the attack model was a Na\"ive Bayes model. Therefore, $type(F_t^{max})$ = DT, $type(F_g^{max})$ = LR, and $type(F_a^{max})$ = NB, as shown in column 2 of Table \ref{tab:highest_acc_combo} along with the attack accuracy under these settings. For the CIFAR-10 dataset, we then also report the attack accuracy when all the models are a decision tree (i.e. equivalent to $type(F_t^{max})$), when all models are logistic regression models ($type(F_g^{max})$), and finally when all models are Bayesian ($type(F_a^{max})$). We repeat this process for all seven datasets. 

We observe that, across datasets,  it is not necessary for all models to be of the same type, as no maximally accurate combination contained the same model type across all three phases of the membership attack development. Additionally, the attack data generation model, as previously assumed, does not need to strictly mirror the target model for a successful membership inference attack. In fact, for 5 out of the 7 datasets the highest accuracy was reported when the attack data generation model was of a different type than the target model, i.e. $type(F_t^{max}) \neq type(F_g^{max})$.
The reason behind this non-intuitive phenomenon lies in a more precise understanding of the role of the generation model. Although it was previously assumed that the role was to mirror the behavior of the target model, 
we assert that the generation model's role is to characterize how the target model may be impacted by the inclusion of a particular instance. That is, how the decision boundary of the target model may reveal the inclusion of an instance.

The generation algorithm is therefore trying to characterize probability distributions related to a decision boundary which either has or has not been informed by the instance. From this perspective, it is now more clear that the vulnerability is more closely related to two elements: distribution of the data and sensitivity of the decision boundary. If a decision boundary for a given dataset is notably impacted by the inclusion of a given instance then membership inference attacks are likely to be more successful.

In summary, we have demonstrated that, despite natural intuition, it is not strictly necessary for the attack data generation technique to be of the same type as the target model for a membership inference attack to be successful. 
Additionally, we again see the dominating factors are the target model type and the training dataset. We note that when all model types are equivalent to $F_t^{max}$ the resulting accuracy is within 0.16\% of the maximum reported accuracy. By comparison, setting all model types to $F_g^{max}$ decreases attack accuracy by up to 28.67\%. Setting all model types to $F_a^{max}$ demonstrates an even larger decrease in many cases as $F_a^{max}$ is reported to be a Na\"ive Bayes model for multiple datasets. As we previously identified the Na\"ive Bayes model as robust against membership inference attack, it is not surprising that attack accuracy will decrease significantly in these cases.

While success is close to maximal in all scenarios when the target model type is known, we re-accentuate the attack success seen in Table \ref{tab:highest_acc_combo} with a mixture of model types. This again supports the conclusion that an adversary need not have this level of insider knowledge to launch a successful attack. Rather, when a model and its training dataset are particularly vulnerable, a variety of attack scenarios are likely to demonstrate success.

\subsection{Impact of Attacker Knowledge}\label{subsec:attackerknow}

Finally, we demonstrate the impact of attacker knowledge on their ability to launch successful membership inference attacks. We evaluate attacker knowledge along two separate vectors: (1) accurate target data and (2) accurate shadow data.

\begin{figure}
    \centering
    \begin{tikzpicture}
    \begin{axis}[
        title={Impact of Noisy Target Data on Attack Accuracy},
        xlabel={Noise Upper Bound},
        ylabel={Accuracy},
        xmin=0, xmax=1,
        ymin=50, ymax=100,
        xtick={0,.2,.4,.6,.8,1.0},
        ytick={50,55,60,65,70,75,80,85,90,95,100},
        legend pos=north east,
        ymajorgrids=true,
        grid style=dashed,
    ]
    \addplot[
        color=blue,
        mark=square,
        ]
        coordinates {
        (0,67.09)
        (0.1,65.37)
        (0.2,63.88)
        (0.3,60.43)
        (0.4,60.48)
        (0.5,58.33)
        (0.6,57.53)
        (0.7,55.97)
        (0.8,55.35)
        (0.9,54.07)
        (1,53.95)
        };
    \addplot[
        color=ForestGreen,
        mark=x,
        ]
        coordinates {
        (0,66.95)
        (0.1,68.72)
        (0.2,66.01)
        (0.3,62.9)
        (0.4,60.07)
        (0.5,58.29)
        (0.6,57.58)
        (0.7,54.94)
        (0.8,54.44)
        (0.9,54.03)
        (1,52.72)
        };
    \addplot[
        color=red,
        mark=triangle,
        ]
        coordinates {
        (0,80.69)
        (0.1,80.4)
        (0.2,77.47)
        (0.3,73.93)
        (0.4,68.23)
        (0.5,64.73)
        (0.6,61.51)
        (0.7,59.78)
        (0.8,58.21)
        (0.9,57.91)
        (1,56.02)
        };
    \addplot[
        color=Yellow,
        mark=*,
        ]
        coordinates {
        (0,88.64)
        (0.1,86.38)
        (0.2,85.85)
        (0.3,84.66)
        (0.4,83.21)
        (0.5,79.12)
        (0.6,73.5)
        (0.7,70.43)
        (0.8,67.16)
        (0.9,65.21)
        (1,62.44)
        };
        \legend{CIFAR-10, Purchases-10, Purchases-20, Purchases-50}
    \end{axis}
    \end{tikzpicture}
    \caption{Impact of Noisy Target Data on Attack Accuracy with Logistic Regression models.}
    \label{fig:noisy_attack}
\end{figure}
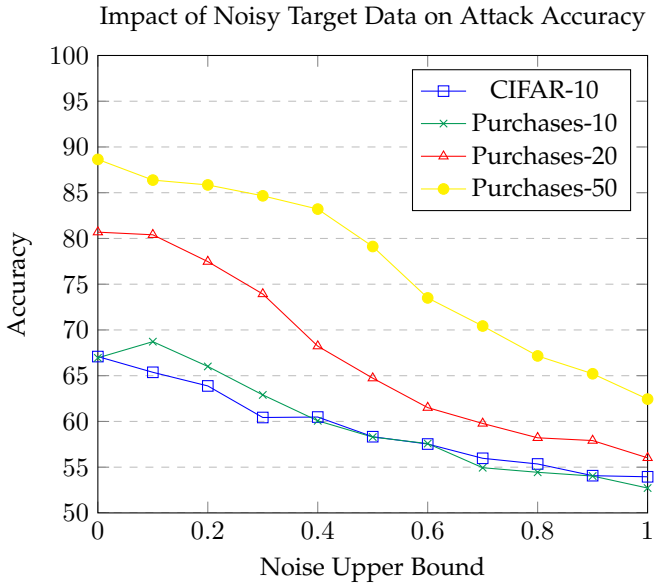

In Figure \ref{fig:noisy_attack}, we demonstrate the impact of attacker knowledge of the target data points. That is, if the attacker does not accurately know each feature within $\mathbf{x}$, is the attacker still able to identify if $\mathbf{x} \in D$ or if $\mathbf{x} \notin D$? To simulate this uncertainty we add noise to the target data points prior to querying. Given features normalized within $[0,1]$ we add noise uniformly sampled from $[0,\sigma]$. In Figure \ref{fig:noisy_attack}, we gradually increase $\sigma$ from 0 to 1, recording the impact on the membership inference attack accuracy. We report results from 4 datasets using logistic regression models for target, generation, and attack models.

While it is unsurprising that the noisier the target instance the less accurate the attack, it is also interesting to note that accurate attacks may still be launched when small amounts of noise are added. That is, if the attacker knows $90\%$ of the data for the target instance but perhaps has to make an educated guess for the final $10\%$, then that attacker is able to identify whether or not the target instance was within the training data with similar accuracy as was seen when $100\%$ of the target data was known.

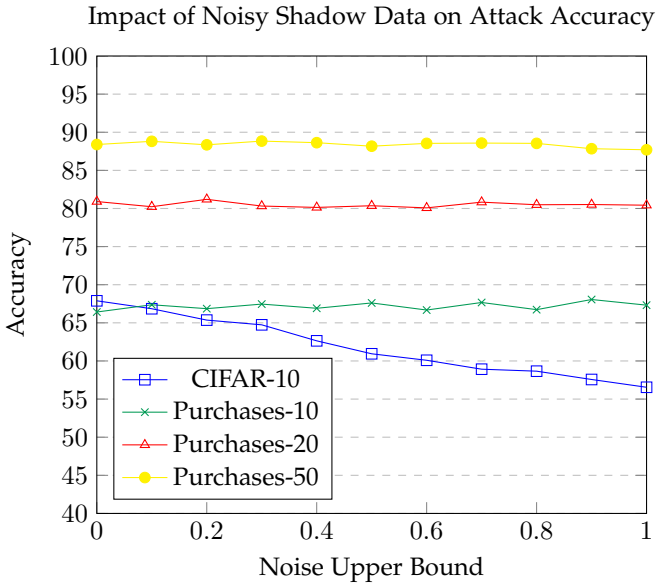
\begin{figure}
    \centering
    \begin{tikzpicture}
    \begin{axis}[
        title={Impact of Noisy Shadow Data on Attack Accuracy},
        xlabel={Noise Upper Bound},
        ylabel={Accuracy},
        xmin=0, xmax=1,
        ymin=40, ymax=100,
        xtick={0,.2,.4,.6,.8,1.0},
        ytick={40,45,50,55,60,65,70,75,80,85,90,95,100},
        legend pos=south west,
        ymajorgrids=true,
        grid style=dashed,
    ]
    \addplot[
        color=blue,
        mark=square,
        ]
        coordinates {
        (0,67.89)
        (0.1,66.85)
        (0.2,65.36)
        (0.3,64.74)
        (0.4,62.64)
        (0.5,60.94)
        (0.6,60.09)
        (0.7,58.92)
        (0.8,58.66)
        (0.9,57.57)
        (1,56.55)
        };
    \addplot[
        color=ForestGreen,
        mark=x,
        ]
        coordinates {
        (0,66.42)
        (0.1,67.37)
        (0.2,66.86)
        (0.3,67.46)
        (0.4,66.91)
        (0.5,67.61)
        (0.6,66.68)
        (0.7,67.67)
        (0.8,66.73)
        (0.9,68.06)
        (1,67.32)
        };
    \addplot[
        color=red,
        mark=triangle,
        ]
        coordinates {
        (0,80.91)
        (0.1,80.23)
        (0.2,81.2)
        (0.3,80.32)
        (0.4,80.14)
        (0.5,80.36)
        (0.6,80.08)
        (0.7,80.83)
        (0.8,80.49)
        (0.9,80.52)
        (1,80.43)
        };
    \addplot[
        color=Yellow,
        mark=*,
        ]
        coordinates {
        (0,88.39)
        (0.1,88.81)
        (0.2,88.35)
        (0.3,88.84)
        (0.4,88.63)
        (0.5,88.17)
        (0.6,88.54)
        (0.7,88.58)
        (0.8,88.54)
        (0.9,87.84)
        (1,87.7)
        };
        \legend{CIFAR-10, Purchases-10, Purchases-20, Purchases-50}
    \end{axis}
    \end{tikzpicture}
    \caption{Impact of Noisy Shadow Data on Attack Accuracy with Linear Regression Models.}
    \label{fig:noisy_shadow}
\end{figure}

We conducted similar experimentation with respect to the shadow data. That is, we began by selecting a shadow dataset disjoint from the target training data but from the same distribution. We then added noise to each instance within the shadow dataset as was done in the previous experiment. Shadow models were then trained and evaluated during the attack data generation process using this noisy shadow dataset. Figure \ref{fig:noisy_shadow} reports the impact of $\sigma$ on the resulting membership inference attack accuracy. Experiments were again run for 4 datasets with the logistic regression model used in all phases.

We can see in Figure \ref{fig:noisy_shadow} that the attack accuracy is relatively resilient to noisy shadow data. While the CIFAR-10 dataset shows a slight decline in accuracy as noise increases, this decline is minor when compared to the inclusion of noise into the target data. Additionally, the various Purchases datasets show strong resilience to the introduction of noise into the shadow dataset. When comparing Figures \ref{fig:noisy_attack} and \ref{fig:noisy_shadow} it is clear that the attacker is more likely to be successful if resources are allocated to developing strong, accurate target instances compared to perfectly representative shadow data. 

\begin{figure}
    \centering
    \begin{tikzpicture}
    \begin{axis}[
        title={Impact Shadow Data Size on Attack Accuracy},
        xlabel={Number of Rows in $D'$},
        ylabel={Accuracy},
        xmin=0, xmax=2500,
        ymin=50, ymax=100,
        xtick={0,500,1000,1500,2000,2500},
        ytick={50,55,60,65,70,75,80,85,90,95,100},
        legend pos=north west,
        ymajorgrids=true,
        grid style=dashed,
    ]
    \addplot[
        color=blue,
        mark=square,
        ]
        coordinates {
        (50,58.89)
        (200,59.11)
        (300,58.52)
        (400,58.44)
        (500,59.155)
        (600,61.26)
        (700,60.65)
        (800,62.34)
        (900,61.04)
        (1000,62.635)
        (1100,63.47)
        (1200,63.9)
        (1300,63.97)
        (1400,62.77)
        (1500,63.72)
        (1600,65.02)
        (1700,64.29)
        (1800,64.74)
        (1900,65.51)
        (2000,65.555)
        (2100,65.81)
        (2200,65.97)
        (2300,64.66)
        (2400,65.45)
        (2500,66.125)
        };
    \addplot[
        color=ForestGreen,
        mark=x,
        ]
        coordinates {
        (51,64.56)
        (200,69.18)
        (300,69.06)
        (400,67.96)
        (500,67.02)
        (600,67.34)
        (700,66)
        (800,66.26)
        (900,65.37)
        (1000,65.39)
        (1100,64.79)
        (1200,65.08)
        (1300,65.46)
        (1400,64.98)
        (1500,65.99)
        (1600,65.35)
        (1700,65.43)
        (1800,65.11)
        (1900,65.19)
        (2000,64.77)
        (2100,64.58)
        (2200,65.07)
        (2300,64.49)
        (2400,64.26)
        (2500,64.31)
        };
    \addplot[
        color=red,
        mark=triangle,
        ]
        coordinates {
        (53,67.58)
        (200,75.9)
        (300,78.58)
        (400,79.41)
        (500,79.64)
        (600,81.06)
        (700,79.95)
        (800,80.35)
        (900,80.43)
        (1000,81.15)
        (1100,80.54)
        (1200,80.45)
        (1300,80.82)
        (1400,80.74)
        (1500,80.31)
        (1600,79.97)
        (1700,80.15)
        (1800,80.54)
        (1900,80.28)
        (2000,79.74)
        (2100,80.74)
        (2200,80.89)
        (2300,80.59)
        (2400,80.91)
        (2500,80.47)
        };
    \addplot[
        color=Yellow,
        mark=*,
        ]
        coordinates {
        (203,72.82)
        (301,77.04)
        (401,77.57)
        (502,78.58)
        (601,79.79)
        (701,80.91)
        (801,80.92)
        (900,81.51)
        (1001,81.32)
        (1100,82.33)
        (1200,83.01)
        (1300,83.5)
        (1400,84.35)
        (1500,84.01)
        (1601,83.37)
        (1700,84.1)
        (1800,85.15)
        (1901,85.99)
        (2000,85.42)
        (2100,85.5)
        (2200,85.81)
        (2300,85.83)
        (2400,86.02)
        (2500,86.53)
        };
        \legend{CIFAR-10, Purchases-10, Purchases-20, Purchases-50}
    \end{axis}
    \end{tikzpicture}
    \caption{Impact Shadow Data Size on Attack Accuracy with Logistic Regression models.}
    \label{fig:num_shadow}
\end{figure}
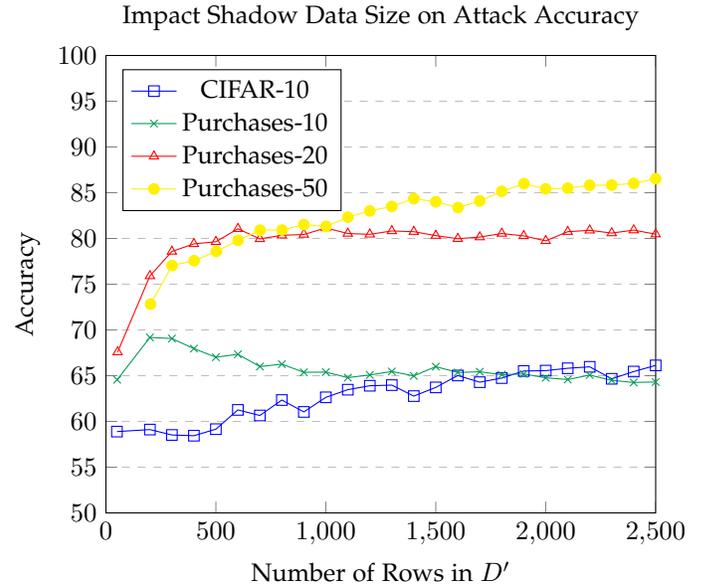

Finally, in Figure \ref{fig:num_shadow} we report the effect that the size of the shadow dataset has on attack accuracy. For these experiments we select shadow datasets of various sizes which are again disjoint from the target training data but from the same distribution. No noise is added for these experiments to isolate the impact of the shadow data size. The experiments are evaluated on the same four datasets again with logistic regression models.

Each dataset in these experiments appears to result in different behavior. The CIFAR-10 dataset continues to show increased accuracy as the shadow dataset size increases. The Purchases-10 dataset however shows an initial accuracy increase before decreasing again and leveling out by the time the shadow dataset reaches 1,000 instances. The Purchases-20 dataset shows a fast initial trend of accuracy improvement before leveling off at a shadow dataset size around 700 instances. Finally, the Purchases-50 dataset results show a similar trend to that seen with the CIFAR-10 dataset where the accuracy continues to increase with the shadow dataset size.

Overall through Figures \ref{fig:noisy_attack}, \ref{fig:noisy_shadow}, and \ref{fig:num_shadow} it is clear that accuracy in the target instances is the most important attacker knowledge factor in determining attack success. With limited resources an attacker should therefore focus on acquiring sufficient information with respect to target instances. The next most deterministic element is the size of the shadow dataset. Rather than allocating resources to ensuring completely accurate instances, attackers should instead put the focus on quantity in the development of their shadow datasets.

\subsection{Federated Learning and Insider Attacks}\label{subsec:insider_intro}

\subsubsection{Insider Attack Model}\label{subsubsec:insider_model}

To this point we have exclusively discussed ``outsider'' membership inference attacks. That is, membership inference attacks which are launched by an adversary who is only a user of the target model through black-box access to the target service prediction API. We now introduce the threat of insider membership inference attacks. We define these attacks to be those launched by a participant in a federated learning system. 

In recent years there has been an increased interest in the role of federated learning systems in addressing privacy concerns in data mining. The intuition behind federated systems to protect against membership inference is as follows: if an adversary is able to identify that a certain instance is contained within the training data of a model and that model is the result of a federated learning system, then any individual participant will have plausible deniability with respect to their individual dataset. However, such federations open the door to a new risk through insider attacks.

The difference between an attack on federated systems and outsider membership inference attacks is that, in federated systems, the training dataset $D$ is divided amongst multiple parties who engage in collaborative learning to provide predictions to the machine learning service. We consider the following loosely federated system: given parties $P_1, P_2, ..., P_\ell$ there exist $\ell$ independent dataset $D_1, D_2, ..., D_\ell$, one belonging to each party. Each party $P_i$ will then train a model $F_{t_i}$ using $D_i$ as the corresponding training data. 

Within this environment, new instances will be evaluated as follows. On input of $\mathbf{x}$, each model $F_{t_i}$ will output a probability vector $\mathbf{p_i} = F_{t_i}(\mathbf{x})$. The individual parties will then share their output either openly with one another or with some aggregation service to compute the final output $\mathbf{p} = \overline{ave}(\mathbf{p_1}, \mathbf{p_2}, ..., \mathbf{p_\ell})$ where $\overline{ave}$ refers to point-wise averaging of the probability vectors. Any outside adversary using this service will only have access to the final probability vector $\mathbf{p}$. Any individual party $P_i$ will therefore have plausible deniability because, if an instance is identified as a member of the training data, the adversary is unlikely to identify which training set specifically.

However we must also consider adversaries who are members of the federated learning systems. That is, the aggregation service or a participating party. Under this scenario the ``insider'' will have access to the individual probability vectors $\mathbf{p_1}, \mathbf{p_2}, ..., \mathbf{p_\ell}$. The insider membership inference attack then becomes: given these probability vectors, is the adversary able to identify which dataset a training instance belongs to. 

We now consider \textit{when} parties may participate in a federated system. Let $P_1$, $P_2$, and $P_3$ represent three candidate parties with training datasets $D_1$, $D_2$, and $D_3$ respectively. Let us assume the extreme case that $D_1$, $D_2$, and $D_3$ are statistically equivalent. Then, the trained models $F_{t_1}$, $F_{t_2}$, and $F_{t_3}$ will approximate the same ideal function $\Gamma$. Given this set up, it is then likely that, on any input $\mathbf{x}$, the outputs $F_{t_1}(\mathbf{x})$, $F_{t_2}(\mathbf{x})$, and $F_{t_3}(\mathbf{x})$ will also be statistically equivalent. Here, there is no accuracy gain for $P_1$, $P_2$, or $P_3$ through collaboration. They are therefore unlikely to be motivated to create a federated learning system.

Alternatively, consider such significantly different $D_1$, $D_2$, and $D_3$ such that $F_{t_1}$, $F_{t_2}$, and $F_{t_3}$ may be considered independent. Let us now assume accuracies of 75\%, 80\%, and 70\% and a majority voting aggregation scheme. Such a federation, on input $\mathbf{x}$, has an 84.5\% chance of accurately classifying $\mathbf{x}$, an accuracy higher than any individual model. Under these conditions, $P_1$, $P_2$, and $P_3$ are much more likely to form a federation.

It is therefore reasonable to assume parties are likely to form federated learning systems when their individual datasets are sufficiently different. Unfortunately, this leads to sufficiently different decision boundaries for different parties. These diverging decision boundaries open the door to effective insider membership inference attacks as an adversary will notice differences in $\mathbf{p_1}$, $\mathbf{p_2}$, ..., $\mathbf{p_\ell}$.


\subsubsection{Insider Attack Risk in Federated Systems}\label{subsubsec:insider_exp}


In Table \ref{tab:insider_acc} we see that even datasets showing resilience to outsider membership inference are vulnerable to an insider membership inference attack. We created a federated system where $n = 3$ for datasets with $k \leq 10$ so that each party has sufficient instances of each class to learn a meaningful decision boundary. Given a federation where $n=3$, any party behaving as an adversary will have an attack precision baseline of $50\%$. This allows for comparison with the outsider inference attacks. Both the Adult and MNIST dataset showed minimal vulnerability in the outsider membership inference attacks and experienced significant jumps in vulnerability in the insider attack scenario while the CIFAR-10 and Purchases-10 datasets show similar precision results, notably outperforming the baseline.

\begin{table}[h]
    \centering
    \begin{tabular}{|c||c|c|}
        \hline
        \multirow{2}{*}{Dataset}    & Outsider Inference    & Insider Inference         \\
                                    & Precision (Accuracy)  & Precision (Accuracy)      \\
        \hline
        \hline
        Adult                       & 55.49 (59.89)         & \textbf{73.26} (69.33)    \\
        MNIST                       & 56.66 (61.75)         & \textbf{68.47} (68.17)    \\
        CIFAR-10                    & 83.94 (90.44)         & 82.02 (82.05)             \\
        Purchases-10                & 73.85 (82.29)         & \textbf{74.42} (74.30)    \\
        \hline
    \end{tabular}\newline
    \caption{\small{Insider inference precision in federated systems with 3 parties. Baseline is 50\%. 
    Model type is set to decision tree.}}
    \label{tab:insider_acc}
\end{table}

\begin{figure*}[h]
\centering
\includegraphics[width=0.75\textwidth]{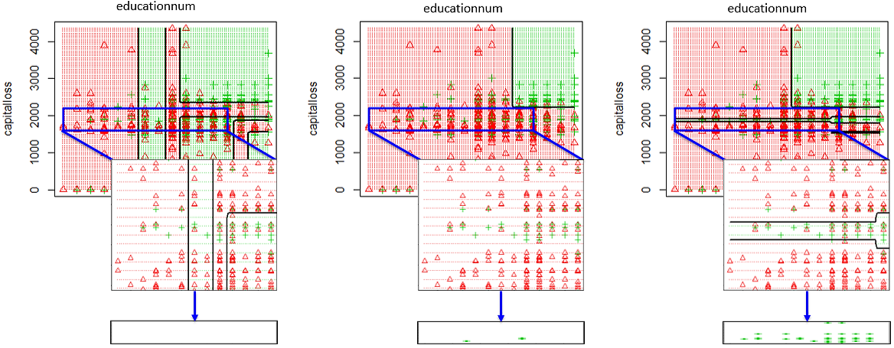}
\caption{\small{Decision boundaries for 3 different participants in a federated system using the Adult dataset. An area where the decision boundaries are significantly different is highlighted as well as the training instances provided to each participant relevant to the highlighted area.}}
\label{fig:owner_boundaries}
\end{figure*}

In Figure \ref{fig:owner_boundaries}, we plot the decision boundaries created by three different decision tree models trained on disjoint subsets of the Adult dataset to articulate how the variation in decision boundaries informs insider attacks. We plot the decision boundaries relative to the \textit{capital loss} and \textit{education number} features. The section enclosed within the blue box highlights a portion of the decision boundary which notably differs between each plot. The second level of Figure \ref{fig:owner_boundaries} is a zoomed-in view of this section for all three plots. On the third level we then plot the positive training instances that informed each decision boundary in this region. It is clear that the long region identified as the positive class in the third plot was informed by significantly more positive instances than the other two decision plots. It is decision boundary differences such as those demonstrated here, and what they reveal of the underlying training data, that reveals ownership in the insider membership inference attack.

More specifically, we know that a participant who returns prediction vectors consistent with the third plot (i.e. higher probabilities for the positive class within the long horizontal region), is more likely to have seen a participant with a capital loss value just under 2,000 and a salary above \$50,000. The probabilities returned from the participant with a model whose decision boundary is more consistent with the first plot however will return higher probabilities for the negative class. It is therefore much less likely that this participant has an individual with a capital loss value within this range and a salary greater than \$50,000. An attacker, therefore, will know that such an individual (i.e. one who falls within this range with a higher salary) is much more likely to be in the dataset of the third participant.

This is supported by the accuracy seen for federations with different data distributions characteristics. When constructing our federated learning systems for these experiments we first sampled target class distributions at random to create different scenarios which may be seen in deployed federated learning environments. 

\begin{figure}[h]
\centering
\includegraphics[width=.75\columnwidth]{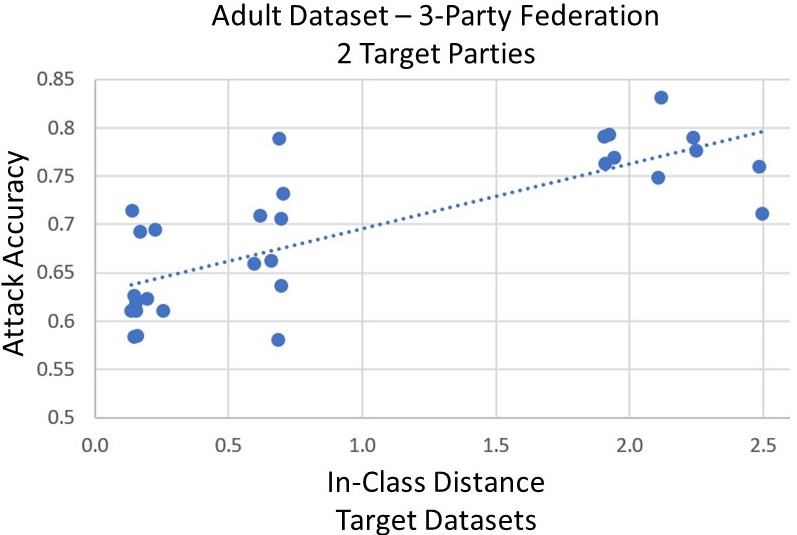}
\caption{\small{In-class distance between different parties compared with insider attack accuracy for the Adult dataset with a 3-party federated learning system.}}
\label{fig:acc_vs_distance}
\end{figure}

In Figure \ref{fig:acc_vs_distance} we show the relationship between the accuracy of the insider membership inference attack and the distance between the two targeted parties' data. That is, let a 3-party federation be formed by parties $P_1$, $P_2$, and $P_3$ wherein $P_3$ behaves maliciously and launches an insider membership inference attack against $P_1$ and $P_2$. Then, we look at how similar all the instances of class $k_i$ in $D_1$ are to those of class $k_i$ in $D_2$. This similarity is averaged across all $k_i \in [1,k]$. If the datasets are more similar then their in-class distance will be lower. In Figure \ref{fig:acc_vs_distance} we can see that $P_3$ will be less successful than when $D_1$ and $D_2$ have a closer in-class measure than if the datasets are very different.

Unfortunately this leads to a catch-22 scenario. Parties with very similar looking training datasets will not be motivated to participate in a federation as they are less likely to see significant increases in classification accuracy. Parties with very different looking training datasets, however, will be more vulnerable to insider membership inference.

We 
argue that the risk of \textit{insider} membership inference attack 
is of particular concern as participants are likely to assume they are less vulnerable than in outsider scenarios due to the plausible deniability protections inherent in federated learning systems. 
The potential for insider attacks, however, calls for a robust trust policy for any federated learning system to consider such risk.


\section{Mitigation Techniques}\label{sec:mitigation}

We categorize mitigation techniques to protect against membership inference into two categories: model hardening and API hardening. 

\textbf{Model Hardening.}  Model hardening mitigation techniques are implemented during the training phase of the target model $F_t$. We suggest four such techniques: (1) model choice, (2) fit control, (3) regularization, and (4) anonymization. In model choice, a service provider may introduce concerns of membership inference into their model selection process. For example, as was demonstrated in our experimentation, a Na\"ive Bayes target model will be much more resilient to membership inference attacks than a decision tree and therefore may be the preferred model type for a particular machine learning service. For fit control, service providers may leverage parameters such as the decision tree's complexity parameter for tree pruning to prevent overfitting and therefore decrease inference risk. Another technique is regularization where noise is added to a model's loss function. This technique is particularly relevant to deep learning models. Finally, the service provider may introduce anonymization techniques into $D$ prior to training. That is, if $D$ is made to be $k$-anonymous prior to the training of $F_t$ then the impact of a single instance may be hid amongst $k$ others prior to training.

\textbf{API Hardening.} API hardening techniques are implemented during the prediction phase through the machine learning service. For example, the service API may introduce noise into the prediction vector before returning it to the user. This will reduce the adversary's understanding of exactly where an instance lies with respect to the target model's decision boundary. Another option is to reduce the dimensionality of $\mathbf{p}$. This can be done either by limiting the return value to the top $k' < k$ values in $\mathbf{p}$ or even returning only the prediction label. Machine learning-as-a-service APIs may additionally be hardened through limiting queries and/or allocating a differential privacy budget to query responses \cite{haeberlen2011differential}. This however requires service providers to track and control user queries which may not always be a suitable option. 

\begin{table}[h]
    \centering
    \resizebox{\columnwidth}{!}{
    \begin{tabular}{|c||c|c|c|l|}
    \hline
        Mitigation                                  & Parameter             & Model Accuracy    & Attack Accuracy   & Utility $\Delta$ \\
        \hline
        \hline
        None                                        &                       & 55\%              & 83\%              & \\
        \hline
        \hline
        \multirow{3}{*}{\shortstack{Dimension\\Reduction~\cite{shokri2017membership}}}
                                                    & $k'=3$                & 55\%              & 83\%              & \multicolumn{1}{|c|}{$\blacktriangledown$} \\
                                                    & $k'=1$                & 55\%              & 82\%              & \multicolumn{1}{|c|}{$\blacktriangledown$} \\
                                                    & $k'=$label            & \textbf{55\%}     & \textbf{73\%}     & \multicolumn{1}{|c|}{$\blacktriangledown$} \\
        \hline
        \hline
        \multirow{4}{*}{Regularization~\cite{shokri2017membership}}
                                                    & L2 $\lambda = 1e-4$   & 56\%              & 80\%              & 
                                                    \hspace{1.5mm}$\blacktriangle$   1\% \\
                                                    & L2 $\lambda = 5e-4$   & 57\%              & 73\%              & 
                                                    \hspace{1.5mm}$\blacktriangle$   2\% \\
                                                    & L2 $\lambda = 1e-3$   & 56\%              & 66\%              & 
                                                    \hspace{1.5mm}$\blacktriangle$   2\% \\
                                                    & L2 $\lambda = 5e-3$   & \textbf{35\%}     & \textbf{52\%}     & \hspace{1.5mm}$\blacktriangledown$  20\% \\
        \hline
        \hline
        \multirow{2}{*}{\shortstack{Adversarial\\Regularization~\cite{nasr2018machine}}}
                                                    & 0                     & 51.9\%            & 63\%              & \hspace{1.5mm}$\blacktriangledown$ 3.1\% \\
                                                    & 2                     & \textbf{47.5}\%   & \textbf{51\%}     & \hspace{1.5mm}$\blacktriangledown$ 7.5\% \\
        \hline
    \end{tabular}}\newline  
    \caption{\small{Results of mitigation techniques using dataset of Texas hospital admissions which contains 100 classes. A neural network target model is used.}}
    \label{tab:mitigation}
\end{table}

An issue that is pervasive in each mitigation technique is a loss of utility. This is demonstrated in Table \ref{tab:mitigation}. We note here that there is not significant reduction in attack accuracy until only the label is returned when using a dimension reduction technique. Consider a hospital processing images of mass scans for cancer classification. A service which says ``This mass is cancerous.'' is significantly less useful than a service which says ``There is a 56\% chance that this mass is cancerous.'' Additionally, even with the strongest dimension reduction, the attack accuracy is still notably outperforming the baseline at 73\%. Regularization on the other hand is able to successfully decrease attack accuracy to 52\%. Unfortunately, to gain this level of protection the noise introduced to the model decreases model accuracy to 35\%. This is a significant challenge in the mitigation of membership inference attacks.

Recent work in \cite{nasr2018machine} has proposed a regularization technique specifically targeting membership inference attacks. While this approach is able to decrease attack accuracy to 51\% with a comparatively moderate decrease in test accuracy to 47.5\%, the training accuracy drops from 81.6\% to 55\%. It is therefore unclear to what degree regularization mitigation is simply a decrease in learning. If the model learns less from the training data in the first place it will naturally have less to reveal when under attack.

\textbf{Differential Privacy.} The definition of membership inference leads to a natural assumption that a solution lies with differential privacy. There are, however, remaining questions on the exact relationship between differential privacy and membership inference attacks as the former is a theoretical framework whereas the latter is an attack with empirical results. Further investigation into this relationship is crucial in understanding what attackers are truly learning in a membership inference attack and to what degree differential privacy curtails attackers' ability to learn this information. 

For example, Figure \ref{fig:noisy_attack} shows that even if an instance which is relatively close to the target instance is provided, the membership inference attack can still be successful. Therefore, is the attacker learning that $\mathbf{x}$ is \textbf{in} $D$, is the attacker learning that something \textit{similar} to $\mathbf{x}$ is in $D$, or is the attacker simply learning that $\mathbf{x}$ is consistent with a decision boundary informed by $D$? 

Additionally, if differential privacy is the answer to mitigating membership inference vulnerability, what level of privacy (value of $\epsilon$) is sufficient to provide protection? For example, the authors in~\cite{rahman2018membership} launch membership inference attacks against deep learning models trained with differentially private mechanisms. While a privacy parameter of $\epsilon = 1$ is able to protect against membership inference, dropping attack accuracy to $50.8\%$ with the CIFAR-10 dataset, the training accuracy also drops from $94.4\%$ in the non-private case to $24.7\%$. Even with a modest privacy parameter of $\epsilon = 8$ the authors report an attack accuracy $58.3\%$ with a training accuracy of just $68.6\%$. The authors in~\cite{pyrgelis2017knock} also remark that defense mechanisms based on differential privacy are not always effective, particularly when an attacker is able to mimic the behavior of the perturbation.

Further research is necessary to understand which differential privacy mechanisms under what settings are able to provide viable protection to membership inference without forfeiting model utility. 

\section{Related Work}\label{sec:related}

Membership inference is a young area, but there still exist a few works since~\cite{shokri2017membership} investigating the risk of membership inference attacks. Most of the existing proposals focus on deep learning models and are influenced by adversarial deep learning research such as~\cite{goodfellow2015laceyella},~\cite{papernot2016limitations},~\cite{yuan2017adversarial}. 
For example,~\cite{long2018understanding} identifies vulnerable instances for membership inference attacks exclusively relating to deep learning models while~\cite{carlini2018secret} seeks to define a measure of deep learning model vulnerability, with respect to the model's encoding of a random secret within the training data, orthogonal to membership inference.~\cite{hayes2017logan} studies membership inference in generative adversarial networks (GANs), and shows
that the level of generalization required to mitigate against membership inference in GANs will lead to worse results in accuracy and utility. This serves as independent evidence for the mitigation strategies and research direction we promote, which include methods for anonymizing the training datasets while preserving the model training quality.

~\cite{long2017towards} proposed a measure of risk at the data instance level, and evaluated the measure on the Adult and Purchases-10 datasets with attack model, target model, and shadow model of the same type. 
The identified instance-level risks exemplify our analysis that membership inference attacks are data-driven.

Alternative study on membership inference relates to the impact of overfitting based on the belief that the cause of membership inference is model overfitting.~\cite{yeom2017unintended} investigates this belief and concludes that overfitting is not necessary for a model to show vulnerability to membership inference. Their investigation is limited to the role of overfitting and assumes a powerful adversary with prior knowledge of the average training loss for $F_t$. This work does motivate that membership inference vulnerability is more complex than just the overfitting in the training data.

Application-specific membership inference, such as ~\cite{pyrgelis2017knock}, studied membership inference vulnerability specific to location data under a powerful adversary with deep prior knowledge. Though this work aims at attacking aggregate data rather than a trained target model and its training data, it does demonstrate the risk of membership inference attacks in a privacy-conscious domain.

Our work is mainly inspired by~\cite{shokri2017membership}, the first exploratory work in membership inference, which shows membership inference risks for a deep learning trained target models. In this paper we extend the work done in~\cite{shokri2017membership} to a more general setting towards demystifying the adverse effect of membership inference across different types of models with both general and empirical characterization of why membership inference attacks are more effective in some scenarios than in others.

\section{Conclusion}\label{sec:conclusion}

We have presented the first generalized framework for the development of a membership inference attack model. This general formulation enables an in-depth characterization of membership infernece attacks against different types of machine learning models. Through extensive experimentation and empirical evidence, we show 
\textit{when} and \textit{why} machine learning models may be vulnerable to membership inference attacks. By exploring a variety of machine learning model types and their correlations with respect to the three phases of the attack generation process, we present five interesting characteristics of membership inference attacks:
(1) they are data-driven attacks, (2) attack models are transferable, (3) target model type is a strong indicator of model vulnerability, (4) attack data generation techniques need not explicitly mirror the target model, and (5) membership inference attacks can persist as insider attacks in federated systems. We also include a discussion on countermeasures and mitigation methods against membership inference attacks.

Our research on membership inference attacks and membership privacy continues along several dimensions. First, we are engaged in the development of countermeasures and defense methods. Second, we are currently studying the scale and diversity of membership inference attacks in federated and collaborative learning systems. Third, we are investigating the complex relationships between membership inference attacks, membership privacy, and differential privacy.


%



\ifCLASSOPTIONcompsoc
  \section*{Acknowledgments}
\else
  \section*{Acknowledgment}
\fi

This research is partially support by the National Science Foundation (NSF) under Grants SaTC 1564097, NSF 1547102, and a Georgia Tech IISP grant. Any opinions, findings, and conclusions or recommendations expressed in this material are those of the author(s) and do not necessarily reflect the views of the NSF and IISP



\ifCLASSOPTIONcaptionsoff
  \newpage
\fi


\input{output.bbl}\bibliographystyle{IEEEtran}

%



%

\begin{IEEEbiography}[{\includegraphics[width=1in,height=1.25in,clip,keepaspectratio]{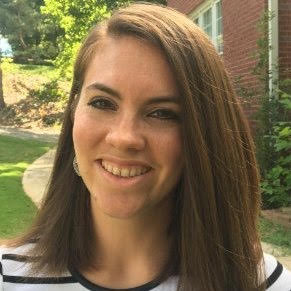}}]{Stacey Truex}
is a PhD student in the School of Computer Science at Georgia Institute of Technology where she is advised by Prof.~Ling Liu. She received her MS degree from the University of Washington in Computer Science and Systems and her BS in Computer Science and BA in Mathematics from Wake Forest University. Her research interests include data privacy, security, federated learning systems, and big data analytics.
\end{IEEEbiography}

\begin{IEEEbiography}[{\includegraphics[width=1in,height=1.25in,clip,keepaspectratio]{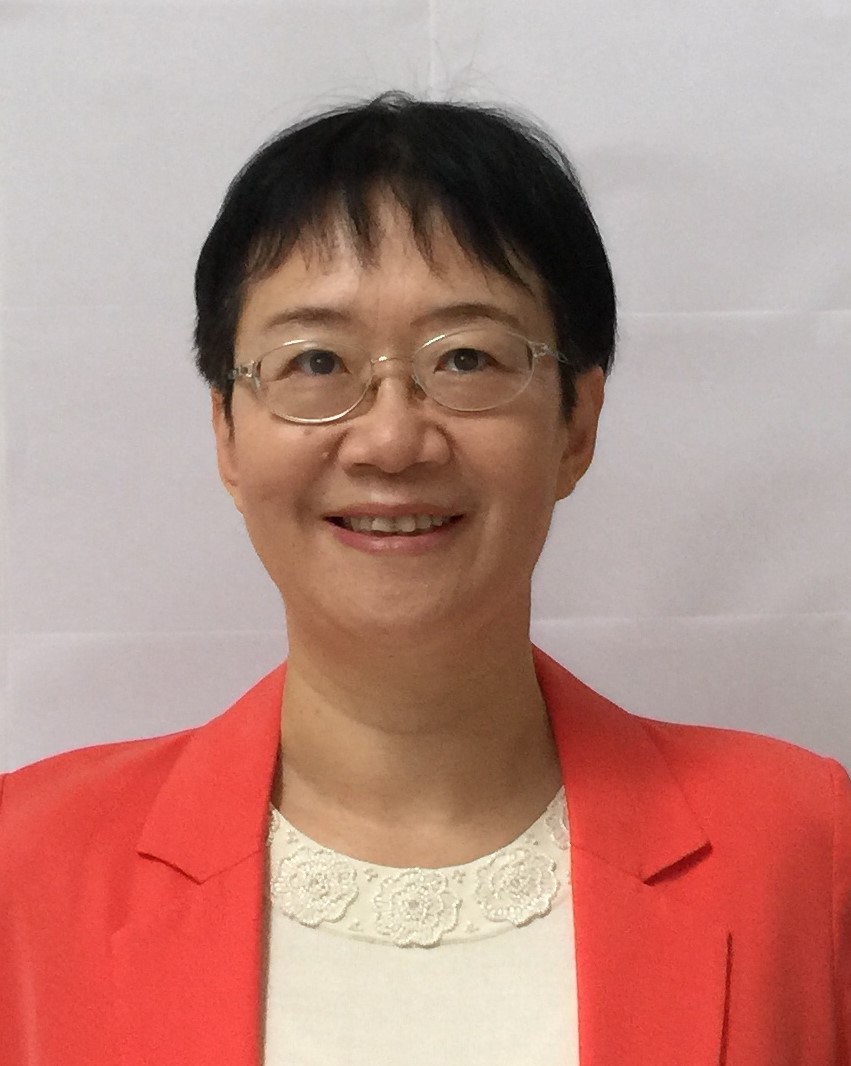}}]{Ling Liu}
is a professor in the School of Computer Science, Georgia Institute of Technology. She directs the research programs in the Distributed Data Intensive Systems Lab (DiSL). She is an elected IEEE fellow, a recipient of the IEEE Computer Society Technical Achievement Award in 2012, and a recipient of the best paper award from a dozen top venues, including ICDCS 2003, WWW 2004, 2005 Pat Goldberg Memorial Best Paper Award, IEEE Cloud 2012, IEEE ICWS 2013, ACM/IEEE CCGrid 2015, and IEEE Symposium on BigData 2016. In addition to serving as the general chair and PC chair of numerous IEEE and ACM conferences in data engineering, she has served on the editorial board of over a dozen international journals. Her current research is primarily sponsored by NSF, IBM, and Intel.
\end{IEEEbiography}


\begin{IEEEbiography}[{\includegraphics[width=1in,height=1.25in,clip,keepaspectratio]{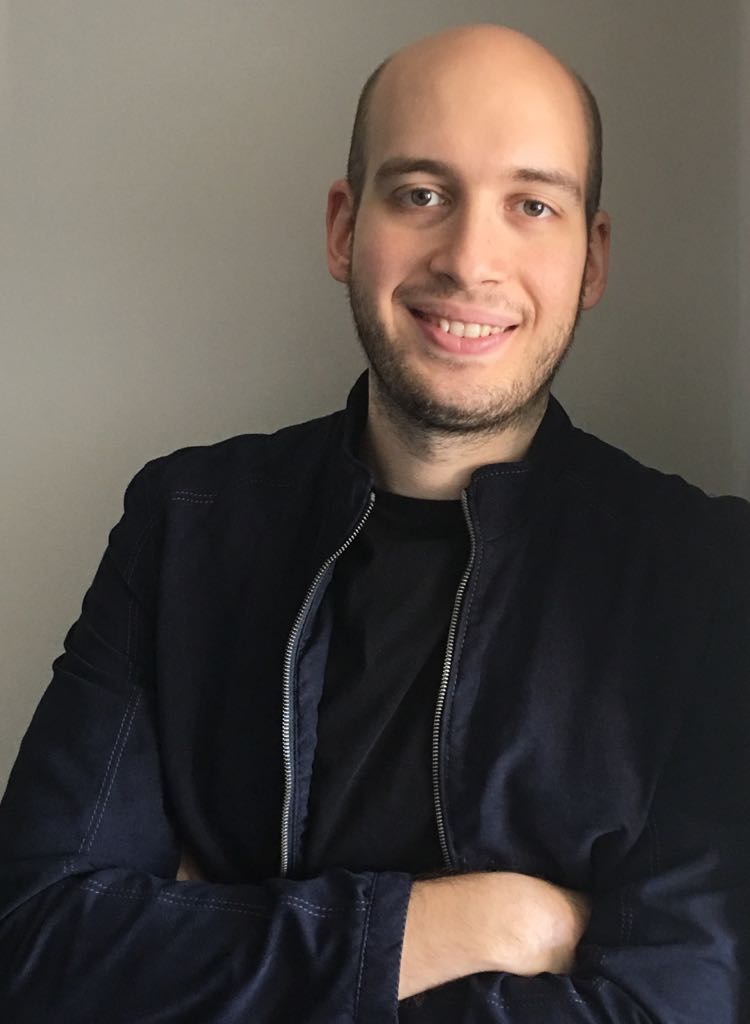}}]{Mehmet Emre Gursoy}
received his MS degree from University of California Los Angeles (UCLA) and his BS degree from Sabanci University, Turkey, both in Computer Science. He is currently pursuing his PhD in the School of Computer Science, Georgia Institute of Technology, where he is advised by Prof.~Ling Liu. His research interests include privacy, security, machine learning, mobile computing, and big data analytics.
\end{IEEEbiography}

\begin{IEEEbiography}[{\includegraphics[width=1in,height=1.25in,clip,keepaspectratio]{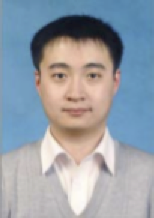}}]{Lei Yu}
is a PhD student in Computer Science in Georgia Institute of Technology. He received the BS and MS degree in Computer Science from Harbin Institute of Technology, China in 2004 and 2006, respectively. His research interests include data privacy, deep learning and big data analytics.
\end{IEEEbiography}

\begin{IEEEbiography}[{\includegraphics[width=1in,height=1.25in,clip,keepaspectratio]{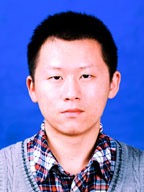}}]{Wenqi Wei}
is currently pursuing his PhD in the School of Computer Science, Georgia Institute of Technology, where he is advised by Prof. Ling Liu. He received his B.E. degree from the School of Electronic Information and Communications, Huazhong University of Science and Technology. His research interests include data privacy, security, machine learning, and big data analytics.
\end{IEEEbiography}




\end{document}